\documentclass[hyper,12pt,a4paper]{JHEP3}
\usepackage{graphicx}
\pdfoutput=1

\usepackage{latexsym,amsmath,amsfonts,amssymb}
\usepackage{mathrsfs}
\usepackage{bbm}
\usepackage{bm}
\usepackage{subfigure}
\usepackage{paralist}
\numberwithin{equation}{section}

\newcommand{\be}{\begin{equation}} 
\newcommand{\ee}{\end{equation}}
\newcommand{\bea}{\begin{equation} \begin{aligned}} \newcommand{\eea}{\end{aligned} \end{equation}}

\newcommand{\bit}{\begin{itemize}} 
\newcommand{\eit}{\end{itemize}}

\newcommand{\cK}{\mathcal{K}}

\newcommand{\cM}{\mathcal{M}}
\newcommand{\cN}{\mathcal{N}}

\newcommand{\cP}{\mathcal{P}}

\newcommand{\cR}{\mathcal{R}}
\newcommand{\cS}{\mathcal{S}}

\newcommand{\cV}{\mathcal{V}}

\newcommand{\cY}{\mathcal{Y}}

\renewcommand{\t}{\widetilde }
\renewcommand{\d}{\partial }
\renewcommand{\b}{\bar}

\newcommand{\dsym}{{\delta_{\text{sym}} }}

\newcommand{\BM}{{\partial \cM}}
\newcommand{\LL}{{\mathscr{L}}}

\title{
On Supersymmetry, Boundary Actions and Brane Charges
}

\author{Lorenzo Di Pietro, Nizan Klinghoffer and Itamar Shamir \\
\it{Weizmann Institute of Science \\ 
Rehovot 76100, Israel}
\email{lorenzo.dipietro@weizmann.ac.il}, 
\email{nizan.klinghoffer@weizmann.ac.il}, 
\email{itamar.shamir@weizmann.ac.il}. 
}

\preprint{WIS/02/15-FEB-DPPA}

\abstract{
Supersymmetry transformations change the Lagrangian $\LL$ into a total derivative $\delta \LL = \partial_\mu \cV^{\mu}$. On manifolds with boundaries the total derivative term is an obstruction to preserving supersymmetry. Such total derivative terms can be canceled by a boundary action without specifying boundary conditions, but only for a subalgebra of supersymmetry. We study compensating boundary actions for $\cN=1$ supersymmetry in 4d, and show that they are determined independently of the details of the theory and of the boundary conditions. Two distinct classes of boundary actions exist, which correspond to preserving either a linear combination of supercharges of opposite chirality (called A-type) or supercharges of opposite chirality independently (B-type). 
The first option preserves a subalgebra isomorphic to $\cN = 1$ in 3d, while the second preserves only a 2d subgroup of the Lorentz symmetry and a subalgebra isomorphic to $\cN = (0,2)$ in 2d. These subalgebras are in one to one correspondence with half-BPS objects: the A-type corresponds to domain walls while the B-type corresponds to strings. We show that integrating the full current algebra and taking into account boundary contributions leads to an energy-momentum tensor which contains the boundary terms. The boundary terms come from the domain wall and string currents in the two respective cases.

}

\begin{document}

\tableofcontents


\section{Introduction}

The problem of preserving supersymmetry on space-time manifolds with boundaries has a long history in the literature. Most notably it has been extensively studied in the context of open strings and D-branes (see for instance \cite{Ooguri:1996ck, Hori:2000ck,  Hori:2000ic, Govindarajan:1999js, Lindstrom:2002mc, Albertsson:2002qc, Hassan:2003uq, Lindstrom:2002jb, Melnikov:2003zv, Koerber:2003ef, Herbst:2008jq}). Much attention was also given to the study of supergravity in spaces with boundaries. This includes applications to the strong coupling limit of $E_8 \times E_8$ heterotic string theory \cite{Horava:1995qa}, supersymmetric Randall-Sundrum models \cite{Altendorfer:2000rr} and general study of supergravity in various dimensions \cite{vanNieuwenhuizen:2005kg,Belyaev:2005rs, Belyaev:2005rt, vanNieuwenhuizen:2006pz,Andrianopoli:2014aqa}.

In field theory, a classification of the half-BPS supersymmetric boundary conditions (BC) for $\cN=4$ Super Yang-Mills was obtained in \cite{Gaiotto:2008sa}, and the behavior of these BC under S-duality was analyzed in a subsequent paper \cite{Gaiotto:2008ak}. With fewer supersymmetries, the general BC and their interplay with dualities are still largely unexplored. (See \cite{Berman:2009kj, Berman:2009xd, Okazaki:2013kaa} for the 3d case.)

Furthermore, recently there has been a great progress, initiated in \cite{Festuccia:2011ws, Dumitrescu:2012ha, Klare:2012gn}, in understanding how supersymmetry can be preserved on curved manifolds. Advances in localization suggest that partition functions factorize on some curved backgrounds, and the factors have the interpretations of partition functions on manifold with boundaries (see for instance \cite{Pasquetti:2011fj, Beem:2012mb, Benini:2012ui}). Motivated by this set of questions, in this paper we consider $\cN = 1$ theories on a 4d space-time with a boundary. 

A supersymmetric Lagrangian $\LL$ transforms under supersymmetry into a total derivative $\delta \LL = \d_\mu \cV^\mu$. When there is a boundary, the variation of the action is a boundary term 
\begin{align}  \label{intro}
\delta \int_\cM \LL = \int_\BM \cV^n~.
\end{align}
Here $\cM$ is space-time, $n^\mu$ is the normal to the boundary and $\cV^n = n_\mu \cV^\mu$. We will consider this as the basic obstruction to preserving supersymmetry. We will show how to construct boundary Lagrangians $\Delta$ for which $\delta \Delta = - \cV^n$ so that 
\begin{align}  \label{}
\delta \left(\int_\cM \LL + \int_\BM \Delta \right) &= 0~.
\end{align}
In this way we can construct actions which are invariant under supersymmetry, independently of the choice of BC. This idea was suggested by several authors \cite{Belyaev:2005rs, Bolle:1989ze, Luckock:1991se, Gates:1997kr, Gates:1998hy, Howe:2011tm, Belyaev:2008xk, Bilal:2011gp}. 
In this paper we explore this idea systematically for $\cN=1$ in 4d. 

Let us demonstrate how this works in an example, given by \cite{Belyaev:2008xk} (see also \cite{Bilal:2011gp}). Consider a superpotential, which comes from a chiral multiplet $W=(w,\psi_w, F_w)$ with supersymmetry variations
\bea  \label{}
&\delta w = \sqrt{2} \zeta \psi_w~, \cr
&\delta F_w = \sqrt{2}i \b \zeta \b \sigma^\mu \d_\mu \psi_w~.
\eea
Clearly $\int_\cM F_w$ is a supersymmetric bulk action. We can use $\Delta = iw$ as a compensating boundary Lagrangian if we restrict to variations for which $\b \zeta_{\dot \alpha} =\zeta^{\alpha}\sigma_{\alpha \dot \alpha}^n $. This defines a subalgebra isomorphic to $\cN=1$ in 3d. It follows that
\begin{align}  \label{}
\int_\cM F_w + i \int_\BM w~
\end{align}
is invariant under this subalgebra without using BC. 

We see in this example that $\cV^n = \sqrt{2} i \b \zeta \b \sigma^n \psi_w$ is exact only with respect to a subalgebra of the supersymmetry transformations. This corresponds to the fact that we cannot preserve all the supersymmetries of the bulk theory. Importantly, we note that the boundary action follows only from the structure of the chiral multiplet. It is independent of the details of the theory and of the specific BC we choose. 

This universality of the boundary action is the first of the two central points of this paper. Focusing on 4d $\cN=1$ it is possible to classify all the $\Delta$'s which solve $-\cV^n = \delta \Delta$ for any supersymmetric Lagrangian. This leads to a classification of the subalgebras that can be preserved in this way. We obtain that they are isomorphic to one of the following
\begin{enumerate}
\item $\cN =1$ in 3d~: by preserving a linear combination of the supercharges $Q_\alpha$ and $\b Q_{\dot \alpha}$ of opposite chirality. This breaks the $R$-symmetry. 
\item $\cN=(0,2)$ in 2d~: by preserving a single component of each chirality independently together with the $R$-symmetry and breaking to 2d Lorentz symmetry on the boundary. 
\item $\cN=(0,1)$ in 2d~: the intersection of the two options above. 
\end{enumerate}
Option (1.) corresponds to the solution of \cite{Belyaev:2008xk, Bilal:2011gp}, while (2.) is to the best of our knowledge novel. 
They are related by dimensional reduction to the familiar $A$- and $B$-type branes in $\mathcal{N}=(2,2)$ in 2d \cite{Ooguri:1996ck, Hori:2000ck}. The third subalgebra is the intersection of the first two. It comes about if we introduce two terms in the boundary action, each preserving only one of the two subalgebras above. In each case, after the boundary action is introduced, one can have various choices for BC which are compatible with the preserved subalgebra. 

Interestingly, the conditions under which the boundary Lagrangians are well-defined operators are exactly the same as the criteria in \cite{Komargodski:2010rb} for the existence of certain supersymmetric multiplets of the energy-momentum tensor. For example, for Abelian gauge theories with a Fayet-Iliopoulos (FI) term the A-type boundary Lagrangian is not gauge invariant. In these theories, the Ferrara-Zumino (FZ) multiplet of the energy-momentum tensor is not defined. Similarly, a theory must have a preserved $R$-symmetry in order to construct the B-type boundary action. Exactly in this case one can define the $\cR$-multiplet of the energy-momentum tensor. Moreover, the subalgebras above are in one-to-one correspondence with those preserved by BPS domain walls (case 1.) strings (2.) or both (3.). In fact, we will see that there is a relation between the boundary Lagrangian and the brane charges appearing in the supersymmetry algebra. These in turn are related to the multiplets of the energy-momentum tensor \cite{Dumitrescu:2011iu}. 

However, it is important to note that the failure of a certain boundary action to exist does not immediately lead to obstructions on preserving the subalgebras above in presence of the boundary. This is because it may be possible to choose appropriate BC that make the operators in the boundary Lagrangian well-defined (we will give an example of how that can happen in the main body). It only represents an obstruction to preserve supersymmetry {\it independently} of the choice of BC.

The relation between a nontrivial $\cV^\mu$ and brane charges in the algebra has been known for a long time \cite{deAzcarraga:1989gm}. 
These brane charges appear in the supersymmetry variation of the supercurrent as follows
\bea  \label{}
\{Q_\alpha, \, \bar{S}_{\dot{\alpha}\nu}\} & = 2 \sigma^\mu_{\alpha \dot{\alpha}} (T_{\mu\nu} + C_{\mu\nu}) +\dots ~, \cr 
\{Q_\alpha, \, S_{\beta\rho} \} & =  \sigma^{\mu\nu}_{\alpha \beta}C_{\rho\mu\nu} + \dots~, 
\eea
where $S_{\beta}^{\mu}$ and $\bar{S}_{\dot{\alpha}}^{\mu}$ are the supercurrents, $C_{\mu \nu}$ is the string current, $C_{\mu \nu \rho}$ is the domain wall current and the ellipses are Schwinger terms. If $\cV^n$ is exact with respect to a certain subalgebra, then all the brane currents which do not respect this subalgebra must drop. For A-type (B-type) the string (domain wall, respectively) current must vanish up to Schwinger terms. The non-vanishing brane current contributes a boundary term when the current algebra is integrated. We will show that it gives exactly the correct boundary action.\footnote{This bears some resemblance to partial supersymmetry breaking in \cite{Hughes:1986dn}. It would be interesting to explore the relation with anomaly inflow and generalized symmetries, see for example \cite{Gaiotto:2014kfa,Dierigl:2014xta}.} 

The interpretation of the boundary actions as arising from brane charges together with the relation to the FZ- and $\cR$-multiplets is the second key point of the paper. It is our hope that this understanding will facilitate the study background supergravity in theories defined on a manifold with a boundary.

The remainder of the paper is organized as follows. In section 2 we will review the basics of symmetries in quantum field theories with a boundary and explain the idea of compensating boundary actions. In section 3 we focus on $\cN=1$ in 4d and explain how to construct boundary actions. In section 4 we show that these results can be interpreted in terms of the brane currents of supersymmetry.

\section{On Symmetries and Boundaries} \label{BC_sym}

In this section we review some basic aspects of theories with boundaries and symmetries. In particular we discuss compensating boundary Lagrangians, and emphasize that they give rise to improvements of certain symmetry currents.

Consider a space-time $\mathcal{M}= \mathbb{R}^{1,2} \times (-\infty,0]$ with a boundary $\BM = \mathbb{R}^{1,2}$. (The convention for the metric is mostly plus.) The boundary is specified by an outward normal vector $n^\mu$ which is normalized to unit length. Only cases in which $n^\mu$ is a constant \emph{space-like} vector are discussed in this paper. We use the index $\mu=0,\ldots,3$ for coordinates in the bulk and $\hat \mu \neq n$ for coordinates on the boundary.\footnote{We use $n$ as an index in the obvious way $x^n = x^\mu n_\mu$.} We will also consider constant time slices, which we denote by $\Sigma$. The index $a \neq 0$ is designated for coordinates on $\Sigma$ and $\hat a \neq 0, n$ for its boundary $\partial \Sigma$ (see figure). 

\begin{figure}[t]
\centering
\includegraphics{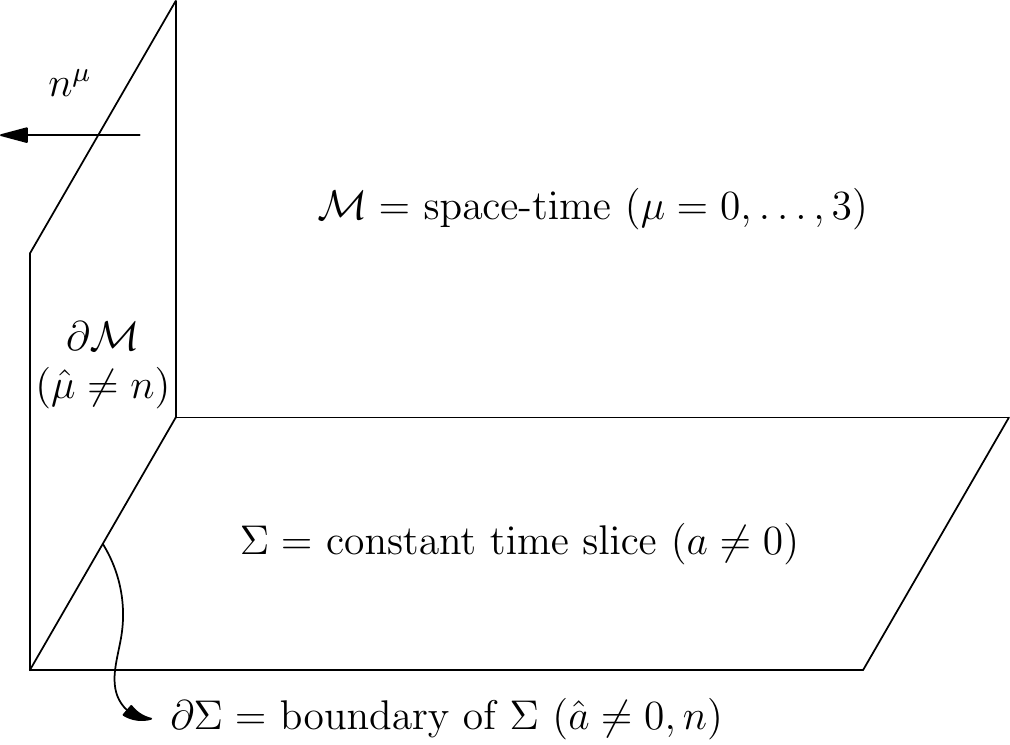}
\end{figure}
The theories we consider are specified by a 4d bulk action $S = \int_{\mathcal{M}} \LL$ and possibly also a 3d boundary action $S_\BM = \int_\BM \LL_\BM$. Taking the variation we get
\begin{align}  \label{var_action}
\delta (S + S_{\partial \cM}) &= \int_{\cM} \left( \frac{\partial \LL}{\partial \Phi} - \partial_{\mu} \frac{\partial \LL}{\partial (\partial_{\mu} \Phi)} \right) \delta \Phi + \int_{\BM} \left(\frac{\partial \LL}{\partial(\partial_{n} \Phi)} \delta \Phi + \delta \LL_{\BM} \right) ~.
\end{align}
Here $\Phi$ represents all the fields in the theory. The BC are relations of the form 
\begin{align} \label{BC1}
\left.G(\Phi, \partial_n \Phi) \right|_{\BM}=0~,
\end{align}
and stationarity of the action on the equations of motion requires that
\begin{align}  \label{BC_equation}
\left.\left(\frac{\partial \LL}{\partial( \partial_{n} \Phi)} \delta \Phi + \delta \LL_{\BM}\right)\right|_{G=0} = \partial_{\hat \mu} (\ldots)~.
\end{align}
For simplicity we are not including additional dynamical fields on the boundary. 

Symmetries are transformations of the fields $\delta_{\text{sym}}\Phi$ which leave the action invariant. (We denote symmetry variations with $\delta_{\mathrm{sym}}$ to distinguish from generic variations $\delta$.) When there are no boundaries, a symmetry is required to satisfy
\begin{align}  \label{bulk_sym}
\delta_{\text{sym}} \LL &=  \partial_{\mu} \cV^\mu~.
\end{align}
By Noether's theorem this implies the existence of a current%
\footnote{
We choose this rather unconventional sign of $J^\mu$ for consistency with the conventions of \cite{Komargodski:2010rb} and \cite{Dumitrescu:2011iu} for the supercurrent and the energy-momentum tensor.} 
\begin{align}  \label{conserved_current}
J^\mu &= - \frac{\partial \LL}{\partial (\partial_{\mu} \Phi)} \delta_{\text{sym}} \Phi + \cV^\mu 
\end{align}
satisfying $\partial_\mu J^\mu =0$ on-shell. This implies that $Q=\int_\Sigma J^0$ is time independent. 
\eqref{conserved_current} is said to be the canonical form of the current. 

Suppose now that there is a boundary, and a transformation satisfying \eqref{bulk_sym} is given. In this context, $\delta_{\text{sym}}$ may be called a \emph{bulk} symmetry. 
Under what conditions does this lead to the implications of a symmetry? Let us mention two aspects of this problem. 

Firstly, the time-derivative of the charge contains a boundary term
\begin{align}  \label{non_const_Q}
\d_0 Q = 
- \int_{\d \Sigma} J^n~.
\end{align}
This means that the conservation may fail because charge can leak through the boundary. The equation $J^n |_{\d \Sigma} =0$ is usually emphasized as the basic requirement of the BC for preserving the symmetry. A different starting point is to demand that the BC \eqref{BC1} are invariant under the symmetry transformation, a criterion that we call symmetric BC. This amounts to imposing\footnote{For space-time symmetries such as supersymmetry which acts with derivatives on fields, we can only demand that \eqref{sym_BC} holds up to equations of motion.} %
\begin{align}  \label{sym_BC}
\delta_{\text{sym}} G |_{G=0} &= 0~.
\end{align}
We will explain below how this leads to the existence of a conserved charge. This condition was discussed by several authors (see for instance \cite{Hori:2000ck,Lindstrom:2002mc,Albertsson:2002qc, Lindstrom:2002jb,Melnikov:2003zv}), mainly as a consistency requirement for supersymmetric BC.

Secondly, in presence of a boundary, a bulk symmetry gives rise to a boundary term in the variation of the action
\begin{align}  \label{obstruction}
\dsym (S + S_\BM ) = \int_{\BM}  \cV^n +  \int_\BM \dsym\LL_\BM ~.
\end{align}
This obstruction to the invariance of the action can be removed \emph{without} invoking the BC (or rather a priori to fixing the BC), by choosing a boundary term which cancels the bulk variation.%
\footnote{Of course, if we assume that the BC are symmetric then it follows from the stationarity of the action \eqref{BC_equation} that the boundary term vanishes on-shell.}

Notably, these two aspects are related because the boundary term that cancels \eqref{obstruction} appears in the stationarity condition \eqref{BC_equation}, in a way which makes it consistent with symmetric BC. Before coming to this point, we proceed to show how symmetric BC lead to vanishing flux. 

\subsection{Constructing Conserved Charges}

It follows from \eqref{non_const_Q} that the charge is conserved if and only if, for our choice of BC, the normal component of the current is a total derivative on $\partial \Sigma$, i.e. 
\bea\label{conscharg}
J^n|_{\partial \Sigma}=\partial_{\hat a} \cK^{\hat a}
\eea
for some $\cK^{\hat a}$ (recall $\hat a \neq 0,n$). Let us show how this is obtained from symmetric BC.

Consider a bulk symmetry as in \eqref{bulk_sym}. 
Using the equations of motion we can write the variation of the bulk action as
\begin{align}  \label{on-shell}
\dsym S |_{\rm on-shell} = \int_\BM \frac{\partial \LL}{\partial(\partial_{n} \Phi)} \dsym \Phi~.
\end{align}
Note that this is valid only if $\dsym$ belongs to the field variations which are allowed by the BC, i.e. we have to consider symmetric BC. Comparing this with \eqref{bulk_sym} we get that 
\begin{align}  \label{}
J^n |_{\partial \cM}= \d_{\hat \mu} \cK^{\hat \mu}~
\end{align}
for some $\cK^{\hat \mu}$ (recall $\hat \mu \neq n$).
This looks very similar to the condition \eqref{conscharg}, but not quite since it includes a time derivative and so does not vanish when integrated on $\partial \Sigma$. This is easily corrected. 

One modifies the definition of the charge by including a boundary term
\begin{align}  \label{}
Q' &= \int_\Sigma J^0 + \int_{\partial \Sigma} \cK^0~.
\end{align}
It is easy to check that $\partial_0 Q' = - \int_{\partial \Sigma} \partial_{\hat a} \cK^{\hat a} =0$. In fact, this can be understood as an improvement of the current. We can find an anti-symmetric tensor $L^{\mu \nu} = - L^{\nu \mu}$ such that $\cK^\mu = L^{\mu n}$. One then constructs an improved current
\begin{align}  \label{}
J'^\mu &= J^\mu + \partial_\nu L^{\mu\nu}
\end{align}
for which $Q' = \int_\Sigma J'^0$ and $J'^n|_{\partial \Sigma} = 0$. Let us stress that it is the canonical current \eqref{conserved_current} which is improved here.

\subsection{Compensating Boundary Lagrangians}

We now turn to a discussion of the boundary terms that can be added to make the action invariant under a symmetry transformation. This idea was first applied to supersymmetry a long time ago \cite{Bolle:1989ze, Luckock:1991se}. More recently it was expounded by Belyaev and van Nieuwenhuizen \cite{Belyaev:2005rs,Belyaev:2008xk,Belyaev:2008ex}. (See also \cite{Berman:2009kj, Howe:2011tm,Bilal:2011gp}.)

Suppose that there exists a $\Delta$ such that
\begin{align}  \label{Delta}
\cV^n + \delta_{\text{sym}} \Delta = \partial^{\hat \mu} \cK_{\hat \mu}~.
\end{align}
In other words $\cV^n$ is exact in the symmetry variation, up to a total derivative on the boundary.  
Then adding $\int_\BM \Delta$ ensures that the action is $\delta_{\text{sym}}$ invariant without reference to BC. We call $\Delta$ a {\it compensating boundary Lagrangian}. 

Beyond the compensating term we have the freedom to add any symmetric boundary action, i.e. a term which is invariant by itself. (Note that $\Delta$ is only defined up to such ``closed'' terms.) This leads to the general form
\begin{align}  
&S  + \int_{\BM} \Delta + \int_{\BM} \LL'_{\BM} ~, \label{total_sym_action} 
\end{align}
where $\delta_{\text{sym}} \LL'_\BM = \partial^{\hat \mu} \cK'_{\hat \mu}$. 

We can use the form of the action in \eqref{total_sym_action} to determine explicitly the required improvement $\cK_{\hat \mu}$ which corresponds to a conserved charge. Let us assume for simplicity that the boundary Lagrangian consists only of the compensating term $\Delta$. Note that equation \eqref{BC_equation} holds with $\delta \to \dsym$ if we have symmetric BC. Then, using \eqref{BC_equation} and \eqref{Delta}, we find%
\footnote{We consider a case where the total derivative in \eqref{BC_equation} vanishes for simplicity.}
\begin{align}  \label{J^n}
J^n|_{\partial \cM} &= -\frac{\partial \LL}{\partial (\partial_{n} \Phi)} \delta_{\text{sym}} \Phi + \cV^n = \d^{\hat \mu} \cK_{\hat \mu}~.
\end{align}
If one introduces in addition a boundary term $\LL'_\BM$ as in \eqref{total_sym_action} then the effect is to change \eqref{J^n} by $\cK_{\hat \mu} \rightarrow \cK_{\hat \mu}+ \cK'_{\hat \mu}$.

Note that, with a compensating boundary Lagrangian, the stationarity condition is manifestly consistent with symmetric BC. This suggests that it is always sufficient to consider actions adhering to the form \eqref{total_sym_action}. One should keep in mind that it is possible to add a boundary term which vanishes trivially on the BC and is not symmetric. The claim is modulo such terms. The mismatch goes also in the other direction. Given an action of the form \eqref{total_sym_action}, it may be possible to choose BC which are not symmetric but still respect the stationarity condition of the action. 

Let us now summarise the discussion above by the following comments. We emphasize that in what follows we will \emph{not} attempt at finding general solutions of $\dsym G |_{G=0} =0$, rather we will focus on equation \eqref{Delta}. The point is that while there are many solutions of $\dsym G =0$, the possible solutions of \eqref{Delta} are finite, each corresponding to a whole family of BC. Moreover, the $\Delta$'s which solve \eqref{Delta} are universal in that they are determined independently of the theory and of the BC. 

\subsection{The Energy-Momentum Tensor}

Let us look closer at the case of translational symmetries, specified by a constant vector $\epsilon^\mu$. 
Since the Lagrangian is a scalar, it follows that $\delta_{\epsilon} \LL = \epsilon^\mu \d_\mu \LL$. 
This gives rise to the canonical energy-momentum tensor
\begin{align}  \label{Tmn_impSec2}
\epsilon^\nu \widehat T_{\nu}{}^\mu &= \epsilon^\nu \left(-\frac{\partial \LL}{\partial (\partial_{\mu} \Phi)} \d_{\nu} \Phi + \delta_{\nu}^{\mu} \LL \right)~.
\end{align}
Here the index $\nu$ is the direction of the translation, and $\mu$ is the current index (i.e. $J_\epsilon^\mu = \epsilon^\nu \widehat T_\nu{}^\mu $) with respect to which it is conserved. In this convention $P_\nu = \int_\Sigma \widehat T_\nu{}^0$ and $P_0 \leq 0$. In general, the canonical energy-momentum tensor is not symmetric. We will use a hat to distinguish it from the symmetric energy-momentum tensor.

Now suppose that there is a boundary. This explicitly breaks translations for which $\epsilon^\mu n_\mu \neq 0$. For the remaining translations with $\epsilon^\mu n_\mu =0$ we have that $\delta_\epsilon S = 0$ and thus they do not require compensating boundary actions. 

Suppose that the definition of the theory includes a boundary Lagrangian $\LL_\BM$.
If a translation by $\epsilon^{\hat \mu}$ is preserved, we must have that $\delta_{\epsilon} \LL_\BM = \partial_{\hat \mu}( \epsilon^{\hat \mu} \LL_\BM) $.
As explained above, this implies an improvement of $\widehat T_\nu{}^\mu$. The precise form depends on $\LL_\BM$, and is necessarily not symmetric (unlike the canonical current which can always be symmetrized). This is linked with the breakdown of Lorentz invariance ensued by the boundary. 

In the discussion above it is important to notice that the energy-momentum tensor that we are improving is the canonical one. This will be important for us because in the context of supersymmetry one usually considers multiplets in which the energy-momentum tensor is symmetric. We will have to take this discrepancy into account.


\section{Boundary Actions in Supersymmetry} \label{SUSY_boundary_actions}

In this section we shall begin our investigation of supersymmetry. The basic constraint on the supercharges that can be preserved in flat space with boundaries arises because supersymmetry transformations anti-commute to translations, some of which are inevitably broken by the boundary. This implies that only a subset of the supersymmetries can survive in presence of boundaries.\footnote{Note that on curved manifolds it is sometimes possible to introduce a boundary without breaking any of the supersymmetries preserved by the background. This is because in general the Killing vectors associated to the isometries that appear on the r.h.s. of the supersymmetry algebra on curved space do not form a basis of the tangent space (see for instance \cite{Dumitrescu:2012ha}). Therefore, one can introduce a boundary that is left invariant by all the isometries appearing in the algebra.}

Focusing on the case of $\mathcal{N}=1$ supersymmetry in 4d, there are two maximal subalgebras that can be preserved, one isomorphic to $\mathcal{N}=1$ in 3d and the other one to $\mathcal{N}=(0,2)$ in 2d. These options correspond to  the possible compensating boundary actions that one can construct. We shall refer to these two cases as A-type and B-type respectively. We find these names appropriate because they are related by dimension reduction to the BC in $\mathcal{N}=(2,2)$ in 2d bearing the same name. Note that in the case of B-type the 3d Lorentz invariance on the boundary is broken by the boundary action.

In $\mathcal{N}=1$ supersymmetry in 4d there are two ways to build bulk actions. One can construct a supersymmetric Lagrangian as the $D$-component of a real multiplet or as the $F$-component ($\b F$-component) of a chiral (anti-chiral) superfield. The basic idea is to use the other bosonic fields in the multiplet to construct compensating boundary terms. We will see below that this follows straightforwardly from the supersymmetry variations which relate the components of the multiplet.  

 We will use the conventions of \cite{Wess:1992cp}, except that we take the Killing spinors $\zeta_\alpha$ and $\b \zeta_{\dot \alpha}$ to be commuting.

\subsection{A-type Boundary Actions}

This is the solution given by Belyaev and van Nieuwenhuizen \cite{Belyaev:2008xk} and later elaborated by Bilal \cite{Bilal:2011gp}, which we now review. (A 2d analogue can be found in \cite{Hori:2000ic,Hori:2013ika}.) In addition, we derive the improvement which follows from the $D$-term action. It will play an important role in section 5.

Let us begin by recalling the example which appeared in the introduction, i.e. the compensating term for the superpotential. The supersymmetric Lagrangian comes from the $F$-component of a chiral multiplet $W=(w,\psi_w,F_w)$. As explained before, it follows from the structure of the chiral multiplet that the boundary term is
\begin{align}  \label{F_obstrct}
\delta_{\b \zeta} \int_{\cM} F_w = \sqrt{2} i \int_{\BM} \b \zeta \b \sigma^n \psi_w~.
\end{align}
To obtain the compensating action we restrict to a subalgebra defined by the relation $\b \zeta_{\dot \alpha} = e^{i \gamma} (\zeta \sigma^n)_{\dot \alpha}$. If the theory has an $R$-symmetry we can set $\gamma=0$ (as assumed in the introduction for simplicity), otherwise it is a free parameter. Equivalently, we consider supercharges which take the form
\begin{align}  \label{typeAcharge}
\t Q_{\alpha} = \frac{1}{\sqrt{2}}  \left( e^{-i\gamma/2} Q_\alpha + e^{i \gamma/2} (\sigma^n \b Q)_\alpha \right)~.
\end{align}
The supersymmetry transformations thus generated are denoted by $\t \delta$. The supercharges satisfy the reality condition
\begin{align} \label{reality}
(\sigma^n \t Q^\dagger)_\alpha = \t Q_\alpha~.
\end{align}

The bulk action supplemented by the boundary term is 
\begin{align}  \label{S_F_Atype}
S_{F,\mathrm{A-type}} = \int_\cM F_w +  i e^{i \gamma} \int_\BM w~.
\end{align}
One can verify that $\t \delta S_{F,\mathrm{A-type}} =0$ with no information assumed about the value of $\psi_w$ on the boundary. Note that the boundary action breaks $R$-symmetry explicitly. 
The subalgebra we obtained is in fact isomorphic to $\mathcal{N}=1$ supersymmetry in 3d
\begin{align}  \label{3d_N=1}
\{ \t Q_\alpha, \t Q_\beta \} = 2 (\Gamma^{ \hat \mu})_{\alpha\beta} P_{\hat\mu}~,
\end{align}
where we defined the 3d gamma matrices by $\Gamma^{\hat\mu} \equiv 2\sigma^{n\hat\mu}$ (recall that $\hat \mu \neq n$), so that $\{ \Gamma^{\hat\mu}, \Gamma^{\hat\nu} \} = -2\eta^{\hat\mu\hat\nu}$. Only momenta tangent to the boundary appear in this algebra. 

We are now ready to consider the $D$-term action. As noted above, the $D$-term resides in a real multiplet whose components are $V=(C,\chi,\b \chi,M,\b M,v_\mu,\lambda,\b \lambda,D)$. They are related by the following transformations
\bea  \label{RealMul}
&\delta C = i \zeta \chi - i \b \zeta \b \chi~, \cr
&\delta \chi_\alpha =  \zeta_\alpha M + (\sigma^\mu \b \zeta)_\alpha (i v_\mu + \partial_\mu C)~, \cr
&\delta \b \chi^{\dot \alpha} =  \b \zeta^{\dot \alpha} \b M + (\b \sigma^\mu \zeta)^{\dot \alpha}(i v_\mu - \partial_\mu C)~, \cr
&\delta M = 2 \b \zeta \b \lambda + 2 i \b \zeta \b \sigma^\mu \partial_\mu \chi~, \cr
&\delta \b M = 2 \zeta \lambda + 2i \zeta \sigma^\mu \partial_\mu \b \chi~, \cr
&\delta v_{\mu} = i \zeta \sigma_\mu \b \lambda + i \b \zeta \b \sigma_\mu \lambda + \partial_\mu (\zeta \chi + \b \zeta \b \chi)~, \cr
&\delta \lambda_\alpha = i \zeta_\alpha D + 2(\sigma^{\mu \nu} \zeta)_\alpha \partial_\mu v_{\nu}~, \cr
&\delta \b \lambda^{\dot \alpha} = -i \b \zeta^{\dot \alpha}D + 2( \b \sigma^{\mu \nu} \b \zeta)^{\dot \alpha} \partial_\mu v_{\nu}~, \cr
&\delta D = - \zeta \sigma^\mu \partial_\mu \b \lambda + \b \zeta  \b \sigma^\mu \partial_\mu \lambda~.
\eea
The top component $D$ is a bulk supersymmetric Lagrangian. Restricting as above to the supercharges $\t Q_\alpha$ we arrive at the following formula for the $D$-term action supplemented by boundary terms
\begin{align}  \label{S_D_Atype}
S_{D,\mathrm{A-type}} = \int_\cM D+ \frac{1}{2} \int_\BM (e^{-i\gamma}M+ e^{i\gamma}\b M) + \int_\BM \partial_n C~.
\end{align}
It is important to note that, unlike the previous case, the boundary terms compensate the bulk variation up to a total derivative on the boundary. 
Explicitly, we have that
\begin{align}  \label{D_B_action}
 \cV^n + \t \delta \left( \frac{e^{-i\gamma}M+ e^{i\gamma}\b M}{2} +  \partial_n C \right) =  i\partial_{\hat \mu} \left(e^{-i\gamma} \b \zeta \b \sigma^{\hat \mu} \chi +  e^{i\gamma}\zeta \sigma^{\hat \mu} \b \chi \right)~.
\end{align}
The significance of this was explained in section 2; a specific improvement of the canonical current is required in order to get a conserved supercharge. Using again the relation $\b \zeta = e^{i\gamma} \zeta \sigma^n$, we find that the improvement of the canonical supercurrent is
\begin{align}  \label{imp_D_Atype}
\zeta \t S^\mu \rightarrow \zeta \t S^\mu -  2i \d_\nu (\zeta \sigma^{\mu\nu}\chi - \b \zeta \b \sigma^{\mu\nu} \b \chi)~.
\end{align}

There is a relation between the $D$-term boundary action and $F$-term boundary action. This comes about because a $D$-term Lagrangian can always be written as a superpotential up to boundary terms. 
More precisely, given a real superfield $V$, we can define a chiral superfield $-\frac{1}{2}\b D^2 V$, whose $F$-component is $D + \d^2 C - i \partial_\mu v^\mu$ and bottom component is $-i\b M$. Using expression \eqref{S_F_Atype} for the $F$-term action with the boundary term, combined with the complex conjugate, leads exactly to the action \eqref{S_D_Atype}. 

\subsection{B-type Boundary Actions}

We have presented above the construction of compensating boundary actions which correspond to the 3d $\mathcal{N}=1$ subalgebra. It is natural to ask if it is possible to preserve supercharges of opposite chirality in an independent way, thus also preserving the $R$-symmetry. Naively the answer to this question appears to be negative: on the boundary we expect to find a supersymmetry algebra with 2 supercharges and the only candidate seems to be the 3d $\mathcal{N}=1$ algebra, whose supercharges are real Majorana fermions and which has no $R$-charge. However, this line of reasoning includes the assumption that 3d Lorentz invariance is maintained. 

Relaxing this assumption, we are allowed to preserve only one component of $Q_\alpha$ and one of $\b Q^{\dot \alpha}$. This is implemented by choosing Killing spinors $\zeta^\alpha$ and $\b \zeta_{\dot \alpha}$.

Without loss of generality we will place $n^\mu$ along one of the axes, by choosing $x^n = x^2$. Let us consider again the $D$-term action. The variations are written as
\begin{alignat}{2}  \label{}
\delta \int_\cM D &= - \int_\BM \zeta \sigma^2 \b \lambda &\qquad \mathrm{and} \qquad 
\b \delta \int_\cM D &= \int_\BM \b \zeta \b \sigma^2 \lambda~.
\end{alignat}
To find compensating boundary actions we choose the Killing spinors $\zeta^\alpha = (1,0)^T$ and $\b \zeta^{\dot \alpha} =(-1,0)^T$, which satisfy the identities $\zeta \sigma^1 = i \zeta \sigma^2$ and $\b \zeta \b \sigma^1 = -i \b \zeta \b \sigma^2$. Using \eqref{RealMul} we then find that a B-type modified $D$-term is given by
\begin{align}  \label{S_D_Btype}
S_{D,\mathrm{B-type}} = \int_\cM D - \int_\BM v_1~.
\end{align}
It is easy to check that this boundary action does not lead to a time derivative on the boundary, so no improvement of the canonical current is needed. The boundary action can also be written as a bulk term $\int_\cM v_{12}$ with $v_{\mu \nu} = \d_\mu v_\nu - \d_\nu v_\mu$. This makes manifest the invariance under shifts of $v_\mu$ by a total derivative.\footnote{One might wonder whether it is possible to preserve two supercharges corresponding to the two components of $\zeta_\alpha$, while breaking $\b \zeta^{\dot \alpha}$ (or viceversa). Indeed one can see from \eqref{RealMul} that an additional possibility for the $D$-term compensating boundary action exists
\begin{align}  \label{}
\int_{\cM} D -i \int_{\BM} v_n~,
\end{align}
which exactly corresponds to preserving only $\zeta_\alpha$. (Preserving $\b \zeta^{\dot \alpha}$ would be achieved by changing the sign of the boundary term.) This subalgebra is not compatible with the requirement that $\b\zeta = \zeta ^\dagger$, which is satisfied in Lorentzian signature, and therefore we reject this possibility. The clash with unitarity is reflected in the boundary action being not real.}

The boundary action explicitly breaks the three dimensional Lorentz invariance by picking a preferred direction $x^1$ on the boundary. We remain with 2d Lorentz invariance in the $(x^0,x^3)$ plane. Defining $Q_- = \zeta^\alpha Q_\alpha$ and $\b Q_- = \b \zeta_{\dot \alpha} \b Q^{\dot \alpha}$ the preserved subalgebra is
\begin{align}  \label{}
\{ Q_-, \b Q_- \} = 2 (P^0 + P^3)~.
\end{align}
This subalgebra is isomorphic to $(0,2)$ supersymmetry in the two dimensions spanned by $x^0$ and $x^3$. Changing the sign of the boundary action in \eqref{S_D_Btype} changes the 2d chirality leading to $(2,0)$ instead.

We now explain how to find B-type compensating boundary action for an $F$-term bulk Lagrangian. To this end, we will see that it is necessary to invoke the existence of an $R$-symmetry. Moreover, differently from all the previous cases, in this case the cancellation of the boundary term will rely on the equations of motion. (It is however independent of the choice of boundary conditions.) For definiteness, we focus on a ($R$-symmetric) superpotential in a Wess-Zumino model.

Consider then a set of chiral fields $\Phi^a$ of $R$-charges $R_a$, a K\"{a}hler potential  $K(\Phi^a, \b \Phi^{\b a})$ and a superpotential $W(\Phi^a)$. The equations of motion are given by
\begin{align}
\b D^2 \partial_a K = 4 \partial_a W~. 
\end{align}
The superpotential must have $R$-charge 2 in order to preserve the $R$-symmetry, i.e. it must satisfy the constraint
\begin{align}
\sum_a R_a \Phi^a \partial_a W = 2 W~.
\end{align} 
Likewise, the $R$-neutrality of the K\"{a}hler potential means that 
\begin{align}  \label{K_R_inv}
\sum_a i R_a \left( \Phi^a \partial_a K - \b \Phi^{\b a} \partial_{\b a} K \right) = 0~
\end{align}
(up to a K\"{a}hler transformation which we disregard for brevity). One can then define a real multiplet $V'=(C',\ldots,D')$ by
\begin{align}  \label{auxRealMul}
V' = \frac{1}{2}\sum_a R_a \Phi^a \partial_a K~.
\end{align}
Using the equations of motion one obtains $\b D^2 V' = 4 W$, which leads to the relation $F_w + \b F_{\b w} =-( D'+ \d^2 C')$. We saw in the study of the $D$-term that the variation of $D'$ is compensated by adding $-v'_1$ on the boundary; $\d^2 C'$ gives rise to an additional boundary term. Hence, we obtain the following form for the $F$-term and the relative compensating boundary Lagrangian
\begin{align}  \label{B_type_F}
S_{F,\mathrm{B-type}} = \int_\cM (F_w+ \b F_{\b w}) + \int_\BM (\partial_n C' + v'_1)~.
\end{align}
To find the corresponding improvement it is useful to note that the fermionic fields of $V'$ and $W$ are related by $\sqrt{2} \psi_w = i \lambda' - \sigma^\mu \d_\mu \b \chi'$. This leads to
\begin{align}  \label{}
\cV^n  = \sqrt{2} i \b \zeta \b \sigma^n \psi_w=- \b \delta \left(\d_n C' + v'_1 \right) - 2i \b \zeta \b \sigma^{n \mu} \d_\mu \b \chi'~,
\end{align}
and similarly for the $\zeta$ variation. We then find that the supercurrents should be improved according to
\begin{align}  \label{imp_F_Btype}
\zeta S^\mu \rightarrow \zeta S^\mu - 2i \zeta \sigma^{\mu \nu} \d_\nu \chi'~, \qquad \qquad
\b \zeta \b S^\mu \rightarrow \b \zeta \b S^\mu + 2i \b \zeta \b \sigma^{\mu \nu} \d_\nu \b \chi'~.
\end{align}

\subsection{Discussion} \label{obstructions}

We would now like to look closer at the boundary actions obtained above, focusing on the cases of a Wess-Zumino model and a $U(1)$ gauge theory. This will expose an intriguing relation to the supersymmetry multiplets of the energy-momentum tensor. Requiring that the boundary actions are well-defined presents nontrivial constraints on the underlying field theory, which will be shown to be equivalent to the existence of those multiplets. 

For a Wess-Zumino model the $D$-term Lagrangian comes from the real superfield $V= \tfrac{1}{2}K$. The A-type compensating boundary Lagrangian for this $D$-term contains the term $\d_n C = \frac{1}{2} \d_n K$. This makes sense only if the K\"{a}hler potential $K$ is well-defined up to an additive constant. Equivalently, the K\"{a}hler connection 
\bea\label{Kform}
-\frac{i}{2}(\d_a K d\Phi^a - \d_{\b a} K \d \b \Phi^{\b a})
\eea
must be globally well-defined. Note that this is never the case if the target space is compact.

Another example comes from the Fayet-Iliopoulos term (FI) in Abelian gauge theories. The real superfield $V$ associated to such a $D$-term action is the elementary Abelian vector superfield. Its bottom component $C$ is shifted by an arbitrary real function under a gauge transformation, making the would-be compensating action $\d_n C$ not gauge invariant. 

On the contrary, the B-type boundary action \eqref{S_D_Btype} for the $D$-term is not affected by any ambiguity in the examples that we have just considered. Both under K\"{a}hler transformations in the Wess-Zumino model, and under gauge transformations in the $U(1)$ gauge theory with an FI term, the boundary Lagrangian changes into a total derivative on the boundary; hence the action is well-defined. On the other hand, we showed that the construction of the B-type boundary actions requires the existence of an $R$-symmetry. (Note that without a superpotential there is always an $R$-symmetry that assigns charge 0 to all the chiral superfields.)

When the boundary Lagrangian does not exist in some theory, it is not possible to obtain a total action that is invariant under the associated subalgebra {\it independently} of the BC. Let us stress that this does not mean that the subalgebra cannot be preserved in this theory. This is because we also need to specify some BC to fully define the theory, and it may be possible that the boundary operator becomes well-defined (or vanish altogether) when evaluated on the BC.

An example will help clarify this issue. Consider a single chiral superfield $\Phi$, whose components we denote by $(\phi, \psi)$, with a canonical K\"{a}hler potential $K = \Phi \b \Phi$. Suppose we identify $\Phi \sim \Phi+1$, i.e. we take the target space to be cylinder. In this case the K\"{a}hler form \eqref{Kform} is \emph{not} globally well-defined, and the term $\frac{1}{2}\d_n (\phi \b \phi)$ in the boundary action is not a well-defined operator. Nevertheless, consider the following BC
 \bea\label{BCcylin}
 \phi & = \b \phi~,\\ \d_n \phi & = - \d_n \b \phi~,\\ \psi & = \sigma^n \b \psi~.
 \eea 
Note that these BC respect the identification $\phi \sim \phi + 1$ on the target space. A short computation reveals that the BC are symmetric with respect to the subalgebra given by the relation $\zeta = \sigma^n \b \zeta$, i.e. an A-type subalgebra. Consistently, note that the boundary action vanishes identically when evaluated on \eqref{BCcylin}. This means that given the BC \eqref{BCcylin} the boundary term is not required for the stationarity of the action, and hence it is redundant. 

Bearing in mind this caveat, we note that the conditions that allow to define the A-type and B-type boundary actions are in one-to-one correspondence with those found by \cite{Komargodski:2010rb} for the existence of the FZ- and $\cR$-multiplets, respectively. These are supersymmetric multiplets of operators that include the energy-momentum tensor. This relation will be elucidated in the next section by a calculation of the current algebra in Wess-Zumino models. In preparation for the next section, let us discuss some relevant aspects of the supercurrent multiplets and their expression in Wess-Zumino models.

The basic fact is that both the FZ-multiplet and the $\cR$-multiplet can only be defined in a restricted class of 4d $\cN=1$ supersymmetric field theories. A third, larger multiplet which exists in general was introduced in \cite{Komargodski:2010rb} and dubbed $\cS$-multiplet. The FZ-multiplet and the $\cR$-multiplet are naturally embedded into the $\cS$-multiplet. When either of the two shorter multiplets is defined, it can be obtained from the $\cS$-multiplet via an improvement transformation (that sets to zero some of its components). A short review of the $\cS$-multiplet and its improvements is given in appendix \ref{S_review}. 

In Wess-Zumino models, given a K\"{a}hler potential $K(\Phi^a, \b \Phi^{\b a})$ and a superpotential $W(\Phi^a)$ the $\cS$-multiplet is given by 
\begin{align}  \label{WessZumino}
&\cS_{\alpha \dot \alpha} = 2 \d_a \d_{\b a}K D_\alpha \Phi^a \b D_{\dot \alpha} \b \Phi^{\b a}~, \cr
&\chi_\alpha = \b D^2 D_\alpha K~, \\
&\cY_\alpha = 4 D_\alpha W~. \nonumber 
\end{align}
Using the improvement \eqref{imprS} we can set $\chi_\alpha=0$ if we choose $U_{FZ}= - \frac{2}{3} K$, and we reduce to the FZ-multiplet. This is an allowed improvement only if the K\"{a}hler potential is well-defined up to an additive constant. On the other hand, to obtain the $\cR$-multiplet we must demand that the theory has an $R$-symmetry. Similarly to the comments in the previous section, when this is the case $U_{\cR} = - \sum_a R_a \Phi^a \d_a K$ is a real multiplet and the equations of motion imply that $\frac{1}{2} D_\alpha \b D^2 U_{\cR} = - 4 D_\alpha W$. The improvement by $U_{\cR}$ sets $\cY_\alpha = 0$ and gives the $\cR$-multiplet, whose bottom component is the conserved $R$-current.

It is interesting to compare the improvements of the supercurrent $S_{\mu\alpha}$ which are implied by the above choices of $U$ with the improvements \eqref{imp_D_Atype} and \eqref{imp_F_Btype} coming from the compensating boundary actions. Consider first the $\cR$-multiplet, compared to the improvement that results from the B-type superpotential. Looking at the $\theta$-component of $U_{\cR}$, we see that the improvement which follows from the $\cR$-multiplet turns out to be twice the B-type improvement \eqref{imp_F_Btype}. We will have to wait until the next section to see how this discrepancy is resolved. It will turn out that the $\cS$-multiplet formulas have to be modified due to boundary effects. 

Now consider the case with well-defined FZ-multiplet and compare to the improvement for the A-type boundary action. We obtained the A-type compensating boundary action for the $D$-term by first rewriting the $D$-term as an integral over only half superspace, and then applying the result for the $F$-term. The resulting $F$-term for a Wess-Zumino model comes from the chiral superfield $4W -\frac{1}{2} \b D^2 K$. Therefore, we have to improve $\cS_{\alpha \dot \alpha}$ in such a way that 
\bea
\cY_\alpha = 4 D_\alpha W \to  D_\alpha(4W -\tfrac{1}{2} \b D^2 K)~.
\eea
This correspond to an improvement with $U'_{FZ} = - K$. (Note that this is different from the improvement $U_{FZ}= - \frac{2}{3} K$ that sets $\chi_\alpha$ to $0$.) Comparing to the improvement that was obtained from the A-type boundary action \eqref{imp_D_Atype}, we find again the same discrepancy by a factor of 2. 


\section{Boundary Actions and Brane Charges} \label{Brane_charges}

In this section we will show that the compensating boundary actions can be interpreted in terms of brane charges of the $\mathcal{N} = 1$ supersymmetry algebra in 4d. From this point of view, a supersymmetric boundary is analogous to a BPS extended object.  The algebra admits two kinds of half-BPS extended objects, namely domain walls and strings, (and quarter-BPS configurations obtained by combining the previous two, i.e. domain wall junctions) \cite{Dvali:1996xe, Chibisov:1997rc,Gorsky:1999hk, Gauntlett:2000ch}. As we will see, they correspond to A-type and B-type compensating boundary actions, respectively. In order to give a self-contained presentation, we will start by briefly reviewing brane charges and BPS objects (see \cite{Shifman:2009zz} for more details). 

\subsection{Brane Charges and BPS Branes in $\mathcal{N} = 1$ in 4d}
The most general $\cN = 1$ supersymmetry algebra in 4d which takes into account brane charges is
\begin{align} 
\{Q_\alpha, \, \bar{Q}_{\dot{\alpha}}\} & = 2 \sigma^\mu_{\alpha \dot{\alpha}} (P_\mu + Z_\mu) \,, \label{algbrane1}\\ \{Q_\alpha, \, Q_\beta\} & =  \sigma^{\mu\nu}_{\alpha \beta}Z_{\mu\nu}~. \label{algbrane2}
\end{align}
The structure of the brane charges $Z_\mu$ and $Z_{\mu\nu}$ is fixed by Lorentz invariance. The real vector $Z_\mu$ is a string charge and the complex two-form $Z_{\mu\nu}$ a domain wall charge. 

The corresponding conserved currents are a two-form current $C_{\mu\nu}$ and a three-form current $C_{\mu\nu\rho}$ which are related to the charges by
\begin{align}
&Z_\mu = \int_\Sigma d^3 x \, {C_\mu}^0~,  \\
&Z_{\mu\nu} = \int_\Sigma d^3 x \, {C_{\mu\nu}}^0~.
\end{align}
In flat space without a boundary the corresponding charge will vanish in any configuration with fields approaching zero sufficiently fast at infinity. This is how one recovers the usual supersymmetry algebra. States carrying brane charges can sometimes be annihilated by a subalgebra of the initial 4d $\mathcal{N} = 1$ supersymmetry algebra. In this case the brane is called BPS.

For instance, for a domain wall with normal vector $n^\mu$, we can go to the rest frame in which $P^\mu = (E, 0,0,0)$, $E$ being the energy of the configuration. The brane charge in this frame can be written as
\begin{align}
Z_{\mu\nu}  = 2 i Z \,  \epsilon_{0 \mu\nu\rho} n^\rho ~, 
\end{align}
where $Z$ is a complex number. $E$ and $Z$ are formally infinite, but the energy and charge per unit volume are finite. 
Consider the supercharges
\begin{align}  \label{}
\t Q_{\alpha} = \frac{1}{\sqrt{2}}  \left( e^{-i\gamma/2} Q_\alpha + e^{i \gamma/2} (\sigma^n \b Q)_\alpha \right)
\end{align}
that appeared in \eqref{typeAcharge}.
Computing their anticommutators in the rest frame, we find
\begin{align}
\{\tilde{Q}_\alpha, \tilde{Q}_\beta\} =   -\Gamma^0_{\alpha \beta} ( 2  E - e^{-i \gamma}Z - e^{ i\gamma} Z^*)= -2  \Gamma^0_{\alpha \beta} (E - |Z|)~.
\end{align}
In the last equality we fixed $\gamma$ to cancel the phase of $Z$. The reality condition of $\tilde{Q}_\alpha$ in \eqref{reality} implies the BPS bound $E \geq |Z|$. When $E = |Z|$, the supercharges $\tilde{Q}_\alpha$ annihilate the state of the domain wall, and the configuration is half-BPS.

Note that, if we consider fluctuations around the state of the domain wall, it is natural to consider a shifted momentum 
\begin{align}
{P'}^{\hat{\mu}} = P^{\hat{\mu}} + |Z| \eta^{\hat{\mu} 0}.
\end{align}
The supercharges $\tilde{Q}_\alpha$ then generate an algebra isomorphic to $\mathcal{N}= 1$ in 3d
\begin{align}
\{\tilde{Q}_\alpha, \tilde{Q}_\beta\} = 2\Gamma^{\hat{\mu}}_{\alpha \beta} P'_{\hat{\mu}}~.
\end{align}

Analogous statements hold for the BPS string associated with the charge $Z_\mu$. In that case we have a real two-form $n^{\mu\nu}$ normal to the two-dimensional world-sheet. In the rest frame the charge can be written as
\begin{align}
Z_\mu = - \tfrac 12 Z \epsilon_{0\mu\nu\rho} n^{\nu\rho}
\end{align}
for a real constant $Z$, and we fixed the normalization so that $n_{\mu\nu}n^{\mu\nu} = 2$ and $Z_\mu Z^\mu=~Z^2 $. We can introduce the chiral projectors
\begin{align}
(\cP_\pm)_\alpha^{~\beta} &= \tfrac 12 (\delta_\alpha^{~\beta} \mp i (\sigma^{\mu\nu})_\alpha^{~\beta}n_{\mu\nu})\,,\\
(\cP_\pm^\dagger)^{\dot{\alpha}}_{~\dot{\beta}} &= \tfrac 12 (\delta^{\dot{\alpha}}_{~\dot{\beta}} \mp i (\bar{\sigma}^{\mu\nu})^{\dot{\alpha}}_{~\dot{\beta}}n_{\mu\nu})~.
\end{align} 
The anticommutator of the projected supercharges ($Q_\pm = \cP_\pm Q $, $\b Q_\pm = \b Q \,\cP^\dagger_\pm $) is
\begin{align}
\{ Q_\pm , \bar{Q}_\pm \} = 2 (E \mp Z)~.
\end{align}
If we take $E = |Z|$, depending on the sign of $Z$ the string will be invariant under the supercharges $+$ or $-$. Shifting the momentum ${P'}^{\hat{\mu}} = P^{\hat{\mu}} + \eta^{\hat{\mu}0} |Z|$, the preserved supercharges will generate an algebra isomorphic to $\cN = (0,2)$ in 2d (or $\cN = (2,0)$ for the opposite sign of $Z$). If both domain walls and strings are present, at most a superalgebra isomorphic to the $\cN = (0,1)$ (or $\cN = (1,0)$) in 2d can be preserved, and the corresponding state is quarter-BPS.

As we have already stressed, the algebras of the supercharges which are symmetries of the BPS domain wall, or the BPS string, are exactly the same algebras which are preserved by the A-type compensating boundary action, or the B-type, respectively. Indeed, we will see in the following subsections that we can interpret such boundary Lagrangians as brane currents supported on the boundary. Taking this point of view, the shift in the momentum ${P'}^{\hat{\mu}} = P^{\hat{\mu}} +\eta^{\hat{\mu}0} |Z|$ reflects the addition of a new term proportional to $|Z|$ to the action (recall that, as discussed in section 2, adding the boundary Lagrangian affects the energy-momentum tensor.) This is the boundary term necessary to obtain an action which is invariant under the preserved algebra. Therefore, this approach will lead to an independent computation of the compensating boundary action, based on the algebra of charges rather than on the variation of the action.

\subsection{Current Algebra of Supersymmetry and Boundaries}

Consider the full current algebra of supersymmetry -- the equal time commutation relations of the supercurrents. Schematically, it takes the form 
\begin{align}  \label{current_alg}
&\{ \b S^0_{\dot \alpha}(t,\mathbf{y}), S^\mu_{\alpha}(t,\mathbf{x}) \} = 2\sigma^\nu_{\alpha \dot \alpha} {T_\nu}^{\mu} \delta^{(3)}(\mathbf{y} - \mathbf{x}) + \dots~, \cr
&\{ S^0_{\alpha}(t,\mathbf{y}), S^\mu_\beta(t,\mathbf{x}) \} = 0 + \dots~, 
\end{align}
where the ellipses represent total derivative terms, usually referred to as Schwinger terms. It is not known in general how to fix the form of all these terms. Note that when there are no boundaries and no extended objects this equation can be straightforwardly integrated to yield the 4d $\cN=1$ supersymmetry algebra $\{ \b Q_{\dot \alpha}, Q_\alpha \}= 2\sigma_{\alpha \dot \alpha}^{\nu} P_\nu$. 

Integrating the anticommutators \eqref{current_alg} only over the $\mathbf{y}$ coordinate on a fixed time slice, one obtains the anticommutator of the supercharge with the supercurrent operator, known as the half-integrated algebra. When there is no boundary, the result of integrating \eqref{current_alg} once is universal for any $\cN=1$ theory in 4d \cite{Komargodski:2010rb}. The following half-integrated current algebra is obtained
\bea  \label{S_mul}
&\{ \b Q_{\dot \alpha} , S_{\alpha \mu}\} = \sigma_{\alpha \dot \alpha}^\nu \left( 2 T_{\nu \mu} + 2 C_{\nu \mu}- \tfrac{1}{2}\epsilon_{\nu \mu \rho \sigma}  \d^\rho j^\sigma+i \d_\nu j_\mu - i \eta_{\nu \mu} \d^\rho j_\rho \right)~, \cr
&\{ Q_\beta, S_{\alpha \rho} \} = \sigma^{\mu \nu}_{\alpha \beta} C_{\rho \mu \nu}~. 
\eea
Here $C_{\mu \nu}$ and $C_{\rho \mu \nu}$ are respectively the string and domain wall currents introduced above. 
Besides the brane currents, an additional operator $j_\mu$ appears in the algebra. The operators in \eqref{S_mul} form the $\cS$-multiplet (reviewed in appendix \ref{S_review}). Let us emphasize that the energy-momentum tensor in \eqref{S_mul} is symmetric. 

As explained in the appendix \ref{S_review}, improvements of the $\cS$-multiplet are parametrized by a real superfield
\begin{align}  \label{Uimpr}
U = u + \theta \eta + \b \theta \b \eta + \theta^2 N + \b \theta^2 \b N - \theta \sigma^\mu \b \theta V_\mu  + \ldots~.
\end{align}
Here we follow the conventions of \cite{Dumitrescu:2011iu}.
This leads to improvements of the energy-momentum tensor and the supercurrent given by
\bea  \label{imp_T_S2}
&S_{\alpha \mu} \rightarrow S_{\alpha \mu} + \d^\nu(2\,\sigma_{\mu \nu}\eta)_\alpha ~, \cr
&T_{\mu \nu} \rightarrow T_{\mu \nu} + \frac{1}{2}(\d_\mu \d_\nu - \eta_{\mu \nu} \d^2)u~.
\eea
Other operators in the $\cS$-multiplet transform as
\begin{align}  \label{imp_others}
& j_\mu \rightarrow j_\mu + V_\mu~, \cr
& C_{\nu \mu} \rightarrow C_{\nu\mu} + \tfrac{3}{4} \epsilon_{\nu\mu\rho\sigma} \d^{\rho} V^\sigma~, \\
&  C_{\nu \mu \rho} \rightarrow C_{\nu\mu\rho} + 2 \epsilon_{\nu\mu\rho\sigma} \d^{\sigma} N~. \nonumber
\end{align}
Note that the improvement preserves the symmetry of the energy-momentum tensor. Under such improvements the half-integrated current algebra \eqref{S_mul} is covariant -- it retains its form when the improvements form a multiplet. In some cases, the improvements can be used to set to zero some of the Schwinger terms. If the brane currents can be improved to $0$, the multiplet is reduced to a shorter one. In particular, when the string current $C_{\mu\nu}$ is set to 0, the shortened multiplet is the FZ-multiplet, while when the domain wall current $C_{\mu\nu\rho}$ is set to 0 we obtain the $\cR$-multiplet.

Consider now the current algebra for theories with a boundary. We wish to integrate \eqref{current_alg} carefully taking into account all the total derivative terms. This will introduce contributions in the integrated algebra of supercharges which have the structure of the brane charges in \eqref{algbrane1} and \eqref{algbrane2}. In analogy with the BPS states, only a subalgebra which is blind to the brane charges can be preserved. Unlike the case with no boundary, the charges are now sensitive to improvements. We must choose the improvements in such a way that the resulting charges are time independent. 

There are several subtleties in realising the idea just presented. Naively, one could just integrate \eqref{S_mul}, with the correct improvement taken into account. However, this does not work for the following reason. The problem is that to obtain \eqref{S_mul} from \eqref{current_alg}, one needs to integrate some total derivative terms in \eqref{current_alg}. We could set their contribution to zero by choosing appropriate BC. However this approach does not allow us to obtain information about the boundary terms. We wish to remain agnostic about a specific choice of BC and keep track of all the boundary contributions.

The following simple example will help explain how boundary terms appear in the algebra, and their relation to the boundary Lagrangian. Consider a real scalar $\varphi$ with a free Lagrangian. The canonical Hamiltonian (density) is given by 
\bea
T_{00} = \frac{1}{2} (\d_0\varphi)^2 + \frac{1}{2} (\d_a \varphi)^2
\eea 
and the canonical commutation relations by 
\bea
i[\d_0 \varphi(\mathbf{x}), \varphi(\mathbf{y}) ] = \delta^{(3)}(\mathbf{x} - \mathbf{y})~.
\eea 
If we pick time-translationally invariant BC, $H = \int_\Sigma T_{00}$ generates the symmetry of time translation. Indeed one readily checks that $i[H, \varphi] = \d_0 \varphi$. However, consider also the action of $H$ on the canonical momenta. Using the equations of motion we obtain
\begin{align}  \label{btermHam1}
i[H, \d_0 \varphi(\mathbf{x})] &= \d_0^2 \varphi(\mathbf{x}) - \delta(x^n) \d_n \varphi(\mathbf{x})~.
\end{align}
Note the additional term localized on the boundary $\{ x^n = 0\}$. Considering the canonical relation 
\bea
i[\d_0 \varphi,\,\cdot \,] = \frac{\delta}{\delta \varphi}( \,\cdot\,)~,
\eea 
we recognize that this boundary term is analogous to the one coming from the variation of the action. Similarly to the latter, also the boundary term in \eqref{btermHam1} must be set to zero by the BC.
In this case it is clear that Neumann BC are implied. We can have Dirichlet BC by adding a boundary term $-\frac{1}{2}\int_{\d \Sigma} \d_n (\varphi^2)$ to $H$, which leads to
\begin{align}  \label{btermHam2}
i[H, \d_0 \varphi(\mathbf{x})] &= \d_0^2 \varphi(\mathbf{x}) + \d_n \delta(x^n)  \varphi(\mathbf{x})~.
\end{align}
Recall that the boundary term affects the Hamiltonian via the improvement of the energy-momentum tensor discussed in section \ref{BC_sym}. We can see from this simple example how the boundary terms in commutation relations with the Hamiltonian are related to boundary terms in the Lagrangian. This will continue to be true for supersymmetry, albeit in a more convoluted way. 

Let us address an objection which might be prompted by the discussion above. We have been using the naive canonical commutation relations, without taking into account how they are modified by the BC. Alternatively, one should first decide on BC and then formulate canonical commutation relations which are consistent with this choice. However, as mentioned before, doing that will prevent us from keeping track of the boundary terms. 

We will bypass this problem in the following way. We consider a theory which is defined in infinite flat space, such that the usual commutation relations hold everywhere. Now we focus our attention on a domain $\Sigma = \{x^n \leq 0\}$ inside the infinite time slice $\mathbb{R}^3$. Charges formed by integration on this restricted domain are of course not guaranteed to be conserved, but there is no problem in computing their commutation relations. In this way we can now use the naive commutation relations and keep track of total derivative terms.

\subsection{Wess-Zumino Model}

In this section we consider a Wess-Zumino model with canonical K\"{a}hler potential and generic superpotential. We will compute the current algebra explicitly starting from the canonical commutation relations, and use it to show the relation between the brane charge and the boundary action. We work in this setup in order to compute explicitly the boundary terms.

Consider a chiral superfield $\Phi=(\phi, \psi,F)$. The K\"{a}hler potential is $K = \b \Phi \Phi$ and the superpotential $W(\Phi)$. The canonical commutation relations are
\begin{align}  \label{commcan}
& i[\d_0 \phi(\mathbf{x}), \b \phi(\mathbf{y}) ] =  \delta^{(3)}(\mathbf{x} - \mathbf{y})~, \\
& \{ \b \psi_{\dot \alpha} (\mathbf{x}), \psi_\alpha (\mathbf{y}) \} = - \sigma^0_{\alpha \dot \alpha} \delta^{(3)}(\mathbf{x} - \mathbf{y})~.
\end{align}
Given the superfield expression $\cS_{\alpha \dot{\alpha}} = 2 D_\alpha \Phi \bar{D}_{\dot{\alpha}} \bar{\Phi}$ for the $\cS$-multiplet, the component operators take the form
\begin{align}\label{componentWZ}
T_{\mu \nu} &= \partial_\mu \b \phi \partial_\nu \phi + \partial_\nu \b \phi \partial_\mu \phi + \eta_{\mu \nu} \left( - \partial_\rho \b \phi \partial^\rho \phi - |w'|  \right) \nonumber \\  & \quad + \tfrac{i}{2}(\b \psi \b \sigma_{(\nu} \partial_{\mu)} \psi + \psi \sigma_{(\nu} \partial_{\mu)} \b \psi)
~,  \\
S_\alpha^\mu &= -\sqrt{2}  \left( (\sigma^\nu \b \sigma^\mu \psi)_\alpha \partial_\nu \b \phi - i (\sigma^\mu \b \psi)_\alpha \b w ' \right)~, \label{SSuperWZ1}\\
\b S^{\dot \alpha \mu} &= -\sqrt{2}  \left( (\b \sigma^\nu \sigma^\mu \b\psi)^{\dot \alpha} \partial_\nu \phi - i (\b\sigma^\mu \psi)^{\dot \alpha} w'  \right)~, \label{SSuperWZ2} \\
C_{\mu \nu \rho} &= -4 \epsilon_{\mu \nu \rho \sigma} \d^{\sigma} \b w~,\\
C_{\mu\nu} &= \tfrac{1}{2}\epsilon_{\mu\nu\rho\sigma}\d^\rho \left(i\phi \,\d^\sigma \b \phi - i\b \phi \, \d^\sigma \phi + \psi \sigma^\sigma \b \psi\right)~, \\
j_\mu &= \psi \sigma_\mu \bar{\psi}~.
\end{align}
As we saw in section \ref{BC_sym}, the Noether procedure gives a different (and non symmetric) expression for the energy-momentum tensor
\begin{align}\label{NoethEM}
\widehat T_{\nu \mu} &= \tfrac{i}{2}\left( \b \psi \b \sigma_\mu \d_\nu \psi + \psi \sigma_\mu \d_\nu \b \psi \right) + \mathrm{bosonic} = T_{\nu \mu} + \tfrac{1}{4} \epsilon_{\nu \mu \rho \sigma} \d^\rho j^\sigma~. 
\end{align}

Writing the supercharge as an integral of the 0-component of the supercurrent, the canonical commutation relations give the expected action on the scalar field%
\footnote{
Note the relation $i (\delta_\zeta+ \delta_{\b \zeta})(\cdot ) = [\zeta^\alpha Q_\alpha + \b \zeta_{\dot \alpha} \b Q^{\dot \alpha}, \, \cdot \,]$~.
}
\begin{align}  \label{}
i[\b Q^{\dot \alpha}, \b \phi(\mathbf{x}) ] = - \sqrt{2}\,  \b \psi^{\dot \alpha}(\mathbf{x}) ~.
\end{align}
On the other hand, we see that an additional boundary term is obtained when acting on the derivative of the field
\begin{align}  \label{}
i[\b Q^{\dot \alpha}, \d_\nu \b \phi(\mathbf{x}) ] = - \sqrt{2}\, \d_\nu \b \psi^{\dot \alpha}(\mathbf{x}) + \sqrt{2} \, \delta^0_\nu \, (\b \sigma^n \sigma^0 \b \psi)^{\dot \alpha} \,\delta(x^n)~.
\end{align}
This entails the following modification of the half-integrated current algebra
\begin{align}  \label{half_int_boundary}
\{ \b Q_{\dot \alpha} , S_{\alpha 0}\} &= \sigma_{\alpha \dot \alpha}^\nu \left( 2 T_{\nu 0} + 2 C_{\nu 0}- \frac{1}{2}\epsilon_{\nu 0 \rho \sigma}  \d^\rho j^\sigma+i \d_\nu j_0 - i \eta_{\nu 0} \d^\rho j_\rho  - i \b \psi \,\b \sigma_0 \sigma_n \b \sigma_\nu \psi \,\delta(x^n) \right)~,
\end{align}
whereas the anticommutators of $Q$ with $S$ and $\bar{Q}$ with $\bar{S}$ are not modified.

Consider now the B-type boundary action, first in the case $W = 0$ in which no improvement is needed to get the conserved supercharge. Recall that in our conventions the normal $n^\mu$ to the boundary is in the direction $x^2$, and the Killing spinors which generate symmetries of the action can be chosen to be $\zeta^\alpha = (1,0)^T$ and $\b \zeta^{\dot \alpha} =(-1,0)^T$. The supercharges $Q_- = \zeta^\alpha Q_\alpha$ and $\b Q_- = \b \zeta_{\dot \alpha} \b Q^{\dot \alpha}$ anticommute to a translation along the light-like Killing vector  $\zeta \sigma^\mu \b \zeta \d_\mu = \d_0 - \d_3$.

Since no improvement is needed in this case, we can obtain the algebra between the conserved supercharges $Q_-$, $\bar{Q}_-$ just by integrating \eqref{half_int_boundary}. Under an integral we can convert the delta function to a total derivative thus leading (with some algebra) to 
\begin{align}  \label{Btype_noF}
\{ \b Q_- , Q_- \} &= \zeta\sigma^\nu\b\zeta \int_\Sigma  \left( 2 T_{\nu 0} + 2 C_{\nu 0} + \frac{1}{2}\epsilon_{\nu 0 \rho \sigma}  \d^\rho j^\sigma \right) \nonumber \\
&= 2\, \zeta \sigma^\nu \b \zeta \int_\Sigma  \left( \widehat T_{\nu 0} + C_{\nu 0}  \right)~.
\end{align}
Note that the additional boundary term $-i \b \psi \,\b \sigma_0 \sigma_n \b \sigma_\nu \psi \,\delta(x^n)$ in \eqref{half_int_boundary} cancels exactly the imaginary term $i \d_\nu j_0 - i \eta_{\nu 0} \d^\rho j_\rho$ upon integration. This is required by consistency, because only hermitean operators can appear on the right hand side of the commutator \eqref{Btype_noF}.

We can express the string current in terms of the components of the real multiplet $V = \tfrac{1}{2} K$. The result is $C_{\nu \mu} = \epsilon_{\nu \mu \rho \sigma} \d^\rho v^{\sigma}$. Plugging this expression, we finally obtain
\begin{align}  \label{}
\{ \b Q_- ,  Q_-\} &= 2\,\left(\int_{\Sigma} \widehat T_{00} - \int_{\Sigma}\widehat T_{3 0} + \int_{\d \Sigma} v_1  \right)= 2(P^0 + P^3)~.
\end{align}
We recognize the boundary term provided by $C_{\nu\mu}$ as the improvement of the canonical energy-momentum tensor \eqref{Tmn_impSec2} associated to the boundary action \eqref{S_D_Btype}. (Recall that our convention is $P_\mu = \int T'^{~ 0}_\mu = - \int T'_{\mu 0}$ where $T'_{\mu \nu}$ is the improved tensor.)

Let us show how improvements of the supercurrents modify the picture. From the multiplet structure of the improvement, encoded in the superfield $U$ in \eqref{Uimpr}, and from the modification $\delta S_{\alpha \mu}=2\,\d^\nu (\sigma_{\mu \nu} \eta)_\alpha$ of the supercurrent, we have 
\bea  \label{S_mul_imp}
&\{ \b Q_{\dot \alpha} , \delta S_{\alpha \mu}\} = \sigma_{\alpha \dot \alpha}^\nu \left( \d_\nu \d_\mu u - \eta_{\nu \mu} \d^2 u + \epsilon_{\nu \mu \rho \sigma}  \d^\rho V^\sigma+i \d_\nu V_\mu - i \eta_{\nu \mu} \d^\rho V_\rho \right) + \ldots~, \cr
&\{ Q_\beta, \delta S_{\alpha \rho} \} = 2 (\sigma^{\mu \nu})_{\alpha \beta} \epsilon_{\rho \mu \nu \sigma} \d^\sigma N~. 
\eea
The ellipsis represent terms localized on the boundary which vanish for the zeroth component, and are therefore not important for us. These terms do not follow from the $\cS$-multiplet and must be verified by an explicit computation. 
The supercharge after the improvement is 
\bea\label{}
Q'_\alpha = \int_\Sigma (S^0_\alpha + \delta S^0_\alpha) = Q_\alpha + \delta Q_\alpha
\eea
where $\delta Q_\alpha$ is a boundary term. Therefore, the total boundary contribution to the algebra due to the improvement is $\{\b Q_{\dot{\alpha}}, \delta{Q}_\alpha  \} + \{\b \delta Q_{\dot{\alpha}}, {Q}_\alpha  \}$.
The net effect is hence twice the real part of the integral of \eqref{S_mul_imp} on the time-slice, and the imaginary part cancels. Similarly, the contribution to $\{ Q_\alpha, Q_\beta \}$ is twice the integral of $\{ Q_\beta, \delta S_{\alpha \rho} \}$ in \eqref{S_mul_imp}. This explains the puzzling factor of 2 that we came across in section \ref{SUSY_boundary_actions}.

Let us now consider the B-type action with $W \neq 0$ which preserves an $R$-symmetry. In this case, we need an improvement \eqref{imp_T_S2}-\eqref{imp_others} with
\bea
U_{\mathrm{B-type}} = - V'  \equiv  -\frac{1}{2} \sum_a R_{a} \b \Phi^{\b a} \d_{\b a} K~,
\eea
where $V'$ is the real multiplet \eqref{auxRealMul} that was used in the construction of the boundary action.
This improvement sets to zero the domain wall current. Indeed, initially we had $C_{\rho \mu \nu} = -4 \epsilon_{\rho \mu \nu \sigma} \d^{\sigma} \b w$, and the effect of the improvement is to change $\b w \rightarrow \b w - N$, where $N$ is the $\theta^2$ component of $U_{\mathrm{B-type}}$. This is the same as the bottom component of
\begin{align}  \label{}
-\frac{1}{4} D^2 \left(-\frac{1}{2} \sum_a R_{a} \b \Phi^{\b a} \d_{\b a} K \right) &= \frac{1}{8} \left( \sum_a R_{a} \b \Phi^{\b a} D^2 \d_{\b a} K \right)  =  \b W~,
\end{align}
and therefore the domain wall current cancels. Computing the change in the boundary terms due to the improvement, as in \eqref{S_mul_imp}, we now get
\begin{align}  \label{}
\{ \b Q'_- , Q'_- \} &= 2\,\left(\int_{\Sigma} \widehat T_{0 0} - \int_{\Sigma}\widehat T_{3 0} + \int_{\d \Sigma} v_1 -  \int_{\d \Sigma} (\d_2C'+ v'_1) \right)~,
\end{align}
where $Q'$ denotes the conserved supercharge. This result is again in agreement with the improvement of the canonical energy-momentum tensor expected from the boundary action \eqref{B_type_F}. In terms of the conserved generators of translations, the algebra is again $\{ \b Q'_- , Q'_-\} = 2(P^0 + P^3)$.

In the computation above we have ignored one term, which we now comment on. This is the anticommutator of the improvements $\{\delta \b Q_{\dot \alpha}, \delta Q_\alpha \}$. Let us evaluate it in a simple example, by considering a canonical K\"{a}hler potential for a single chiral superfield of $R$-charge 1 and vanishing superpotential. It follows that $\delta S^{\alpha\mu} = -\sqrt{2} \d_\nu (\b \phi \psi \sigma^{\nu \mu})^{\alpha}$. We find
\begin{align}  \label{singularity}
\{ \delta \b Q^{\dot \alpha}, \delta S^{\alpha 0}(\mathbf{x}) \} &= \b \sigma^n \sigma^{\mu 0} \d_\mu \left(\phi \b \phi \, \delta(x^n) \right)~.
\end{align}
Integrating once more this term leads to a $\delta(0)$ term on the boundary, which requires some regularization procedure. Since the problem clearly comes from the fact that the boundaries of the two domains of integration coalesce, a simple way to regulate this expression is by slightly changing the range of integration for one of the charges. We keep $\Sigma = \{x^n \leq 0\}$ as before and define another domain by $\Sigma_\varepsilon = \{ x^n \leq  \varepsilon\}$, where $\varepsilon > 0$. We now define $\delta \b Q^{\dot \alpha} $ by integration of $\Sigma_\varepsilon$ instead of $\Sigma$. This changes $\delta(x^n)$ to $\delta(x^n - \varepsilon)$ in \eqref{singularity}. When we now perform the second integration on the domain $\Sigma$, the delta function is not localized on the boundary and the expression vanishes, so that we recover the previous results in the limit $\varepsilon \rightarrow 0$. 

Finally, consider the A-type boundary action. The improvement that we must consider in this case is 
\bea
U_{\mathrm{A-type}} = - V \equiv - \frac 12 K~,
\eea 
where $V$ is again the real multiplet which gives us the $D$-term action.%
\footnote{Note that the component expansion of $V$ is different from that of $U$:
\begin{align}  \label{}
V &= C + i\theta \chi - i \b \theta \b\chi + \frac{i}{2} \theta^2 M - \frac{i}{2} \b \theta^2 \b M - \theta \sigma^\mu \b \theta v_\mu \cr
&\quad + i \theta^2 \b \theta \left( \b \lambda + \frac{i}{2} \b \sigma^\mu \d_\mu \chi \right) - i \b \theta^2 \theta \left( \lambda + \frac{i}{2} \sigma^\mu \d_\mu \b \chi \right) 
 + \frac{1}{2} \theta^2 \b \theta^2 \left(D+ \frac{1}{2}\d^2 C \right)~. 
\end{align}
It is defined in this way so that the variations take the form of \eqref{RealMul}.}
In this case we preserve Majorana supercharges defined by the linear combination 
\begin{align}  \label{}
\t Q_\alpha = \frac{1}{\sqrt{2}} (e^{-i\gamma/2}Q' + e^{i\gamma/2} \sigma^n \b Q')_\alpha~,
\end{align}
where $'$ denotes that we include the boundary contribution from the improvement. We want to compute $\{ \t Q_\alpha, \t Q_\beta \}$ starting from the original current algebra. We take $\gamma=0$ to avoid cluttering, but it should be obvious how this parameter can be reintroduced. 

The anticommutators can be expanded as
\begin{align}  \label{expan}
\{ \t Q_\alpha, \t Q^\beta \} &= \frac{1}{2}\left(\{Q'_{\alpha}, Q^{'\beta}\} - \sigma^n_{\alpha \dot \alpha}  \{\b Q^{'\dot \alpha} , \b Q'_{\dot \beta}\} \b \sigma^{n \dot\beta  \beta} - \{ Q'_\alpha, \b Q'_{\dot \beta} \} \b \sigma^{n \dot \beta \beta} + \sigma^n_{\alpha \dot \alpha} \{ \b Q^{' \dot \alpha} , Q^{'\beta} \}\right)~.
\end{align}
Let us start with the first term on the r.h.s. The effect of the improvement \eqref{imp_others} on the domain wall current is to shift $\b w \rightarrow \b w + \frac{i}{2} M$. Therefore
\begin{align}  \label{}
\{Q'_{\alpha}, Q^{'\beta} \} = 8  \delta^0_{\nu}{(\sigma^{\nu \mu})_\alpha}^\beta \int_\Sigma  \, \d_\mu \left( i \b w  - \frac{1}{2} M \right)~.
\end{align}
Note again that this is twice of the contribution that one would naively expect from the improvement of the $\cS$-multiplet.

Next we consider the opposite chirality commutation relations. It is easy to check that the component $V^\sigma = - v^\sigma$ of $V$ exactly cancels the string current $C_{\nu \mu} = \epsilon_{\nu \mu \rho \sigma} \d^\rho v^\sigma$. The computation gives
\begin{align}  \label{}
\{ \b Q^{' \dot \alpha} , Q^{' \beta} \} = -2\, \b \sigma^{\nu\dot \alpha \beta} \int_\Sigma  \, \left( \widehat T_{\nu 0} - (\d_\nu \d_0 - \eta_{\nu 0} \d^2)C \right)~.
\end{align}
Summing up all the contributions, we obtain
\begin{align}  \label{}
\{ \t Q_\alpha, \t Q^{\beta} \} &= 4 {(\sigma^{n \nu})_\alpha}^\beta \left( \int_\Sigma \widehat{T}_{\nu}{}^{ 0} + \delta^0_{\nu} \int_{\d \Sigma} \left\{ i (w - \b w) + \frac{1}{2} (M+ \b M) +  \d_n C \right\} \right)~.
\end{align}
We find again agreement with the boundary action \eqref{S_F_Atype}-\eqref{S_D_Atype}.

We have thus demonstrated by an explicit computation of the current algebra of the supercharges, for both A-type and B-type subalgebras, how to recover the results of section 3 in an independent way. We obtain the preserved subalgebra with a Hamiltonian modified by boundary terms which exactly correspond to the compensating boundary actions. It is seen explicitly that they are obtained from the brane charges. This establishes the relation between boundary actions and brane charges.


%



\section*{Acknowledgments}

We are grateful to M. Bertolini, C. Closset, J. Gomis, G. Gur-Ari, Y. Oz, F. Porri, C. Sonnenschein, S. Yankielowicz, and especially O. Aharony and Z. Komargodski for useful discussions. 
I.S. is grateful to the Perimeter Institute for Theoretical Physics and the Simons Center for Geometry and Physics, Stony Brook University for their hospitality while some of the research for this paper was conducted. 
The work of L.D.P., N.K. and I.S. was supported by an Israel Science Foundation center for excellence grant (grant no. 1989/14), by the Minerva foundation with funding from the Federal German Ministry for Education and Research, by the I-CORE program of the Planning and Budgeting Committee and the Israel Science Foundation (grant number 1937/12) and by the ISF within the ISF-UGC joint research program framework (grant no. 1200/14).
L.D.P. would also like to thank the United States-Israel Binational Science Foundation (BSF) for support under grant number 2010/629. 
Any opinions, findings, and conclusions or recommendations expressed in this material are those of the authors and do not necessarily reflect the views of the funding agencies.
~

\newpage



\appendix

\section{Multiplets of the Energy-Momentum Tensor}\label{S_review}

The algebra of charges \eqref{algbrane1}-\eqref{algbrane2} can be rewritten in terms of the associated conserved currents as
\begin{align}
\{Q_\alpha, \, \bar{S}_{\dot{\alpha}\nu}\} & = 2 \sigma^\mu_{\alpha \dot{\alpha}} (T_{\mu\nu} + C_{\mu\nu}) +\dots \,, \label{unalgbrane1}\\ \{Q_\alpha, \, S_{\beta\rho} \} & =  \sigma^{\mu\nu}_{\alpha \beta}C_{\rho\mu\nu} + \dots~, \label{unalgbrane2}
\end{align}
where the dots denote possible additional Schwinger terms (i.e. terms compatible with the conservation that do not contribute to the charges). 
These anti-commutation relations imply that the brane currents are related by supersymmetry transformations to the supercurrent $S_{\mu\alpha}$, and therefore they belong to the same multiplet of local operators, which also includes the energy-momentum tensor. Let us provide a brief review of this multiplet following the notation of \cite{Dumitrescu:2011iu}.

Every local $\mathcal{N}=1$ supersymmetric field theory in 4d contains a so-called $\cS$-multiplet \cite{Komargodski:2010rb}, which is a real superfield $\mathcal{S}_{\alpha\dot{\alpha}}$ satisfying
\begin{align}\label{super conservation}
\bar{D}^{\dot{\alpha}} \mathcal{S}_{\alpha \dot{\alpha}} = \mathcal{Y}_\alpha + \chi_\alpha~,
\end{align}
with $\mathcal{Y}_\alpha$ and $\chi_\alpha$ obeying the constraints
\bea
D_\beta \mathcal{Y}_\alpha + D_\alpha \mathcal{Y}_\beta & = 0\,,\quad \bar{D}^2 \mathcal{Y}_\alpha  = 0 \,, \cr
D^\alpha \chi_\alpha - \bar{D}_{\dot{\alpha}} \bar{\chi}^{\dot{\alpha}} & = 0\,,\quad \bar{D}_{\dot{\alpha}} \chi_\alpha   = 0~.
\eea
Solving the constraints one finds the following expansion in components ($\mathcal{S}_\mu \equiv  \frac{1}{4} \bar{\sigma}_\mu^{\dot{\alpha}\alpha} \mathcal{S}_{\alpha \dot{\alpha}}$)
\begin{align}  \label{SM-components}
\cS_\mu &= j_\mu -i \theta \left( S_\mu - \frac{i}{\sqrt{2}} \sigma_\mu \b \psi \right) + i \bar \theta \left( \b S_\mu - \frac{i}{\sqrt{2}} \b \sigma_\mu \psi \right) + \frac{i}{2} \theta^2 \b Y_\mu - \frac{i}{2} \b \theta^2 Y_\mu \nonumber \\
&\quad +(\theta \sigma^\nu \b \theta) \left( 2 T_{\nu \mu} - \eta_{\nu \mu} A - \frac{1}{8} \epsilon_{\nu \mu \rho \sigma} F^{\rho \sigma} - \frac{1}{2} \epsilon_{\nu \mu \rho \sigma} \partial^\rho j^\sigma \right) \nonumber \\
&\quad -\frac{1}{2} \theta^2 \b \theta \left( \b \sigma^\nu \partial_\nu S_\mu + \frac{i}{\sqrt{2}} \b \sigma_\mu \sigma^\nu \partial_\nu \b \psi \right)
+\frac{1}{2} \b \theta^2 \theta \left( \sigma^\nu \partial_\nu \b S_\mu + \frac{i}{\sqrt{2}} \sigma_\mu \b \sigma^\nu \partial_\nu \psi \right) \nonumber \\
&\quad + \frac{1}{2} \theta^2 \b \theta^2 \left( \partial_\mu \partial^\nu j_\nu - \frac{1}{2} \partial^2 j_\mu \right)~, \\
\cY_\alpha &= \sqrt{2} \psi_\alpha + 2\theta_\alpha F + 2i (\sigma^\mu \b \theta)_\alpha Y_\mu -2 \sqrt{2} i \theta \sigma^\mu \b \theta {(\sigma_{\mu \nu})_\alpha}^\beta \partial^\nu \psi_\beta \nonumber \\
&\quad + i \theta^2 (\sigma^\mu \b \theta)_{\alpha} \partial_\mu F + \b \theta^2 \theta_\alpha \partial^\mu Y_\mu - \frac{1}{2 \sqrt{2}} \theta^2 \b \theta^2 \partial^2 \psi_\alpha~,\\
\chi_\alpha &= - i \lambda_\alpha(y) + \theta_\beta \left( \delta^\beta_\alpha D(y) - i {(\sigma^{\mu \nu})_\alpha}^\beta F_{\mu \nu}(y) \right) + \theta^2 \sigma_{\alpha \dot \alpha}^\mu \partial_\mu \b \lambda^{\dot \alpha}(y)~.
\end{align}
Here $T_{\mu\nu}$ is symmetric and $F_{\mu \nu}$ is antisymmetric. These components are not all independent, in order to solve (\ref{super conservation}) we further need to impose
\bea
\partial^\mu T_{\mu\nu}  = 
\partial^\mu S_{\mu\alpha}  =
\partial_{[\mu} Y_{\nu]} =
\partial_{[\mu} F_{\nu\rho]}   & = 0 \cr
\eea
and
\bea
& 4 T^\mu_\mu = 6 A - D ~, \cr
&i \partial^\mu j_\mu = F - A ~, \cr
& 2(\sigma^\mu\b S_\mu)_\alpha = \lambda_\alpha -  3\sqrt{2} i \psi_\alpha~.
\eea
Taking into account these relations, the multiplet contains 16 + 16 independent degrees of freedom (bosonic + fermionic). Note that $T_{\mu\nu}$ is the symmetric and conserved energy-momentum tensor, and $S_{\mu\alpha}$ is the conserved supercurrent.

The supersymmetric variations of the supercurrent operator are given by
\bea \label{varbarS}
i \delta \bar{S}^\mu_{\dot{\alpha}} \equiv \zeta^\beta\{ Q_\beta, \bar{S}^\mu_{\dot{\alpha}} \}& =  (\zeta \sigma_\nu)_{\dot{\alpha}} \left[2 T^{\nu\mu} - i \eta^{\nu\mu} \partial^\rho j_\rho + i \partial^\nu j^\mu - \tfrac 12\epsilon^{\nu \mu\rho\sigma}(\partial_\rho j_\sigma + \tfrac 14 F_{\rho\sigma})\right]~,  \cr
i \delta S^\mu_\alpha \equiv \zeta^\beta \{ Q_\beta, S^\mu_\alpha \}& =  2i ( \zeta \sigma^{\mu\nu}  )_\alpha \bar{Y}_\nu~.
\eea
Comparing with (\ref{unalgbrane1}-\ref{unalgbrane2}) we can identify the brane currents as 
\begin{align}
&C_{\mu\nu} = -\tfrac 1{16} \epsilon_{\mu\nu\rho\sigma} F^{\rho\sigma}~, \label{stringcur}\\
&C_{\mu\nu\rho} = - \epsilon_{\mu\nu\rho\sigma} \bar{Y}^\sigma~.\label{DWcurr}
\end{align}
Note that indeed the additional term in (\ref{varbarS}) besides $T_{\mu\nu} + C_{\mu\nu}$ is a Schwinger term.

The conserved current operators are not defined univocally, they can be changed by improvement transformations. For the $\cS$-multiplet the possible improvements are given in terms of a real superfield $U$, and take the form
\begin{align}
\cS_{\alpha\dot{\alpha}} & \to \cS_{\alpha\dot{\alpha}} + [D_\alpha, \b D_{\dot{\alpha}}] U~, \nonumber \\ 
\cY_\alpha & \to \cY_\alpha + \frac 12 D_\alpha \bar{D}^2 U ~, \label{imprS} \\ 
\chi_\alpha & \to \chi_\alpha + \frac 32 \bar{D}^2 D_\alpha U~. \nonumber
\end{align}

In some cases it is possible to reduce the $\cS$-multiplet to a smaller multiplet. This happens when $\chi_\alpha$ or $\cY_\alpha$ can be set to zero by an improvement transformation \eqref{imprS}. When $\chi_\alpha$ can be set to zero, the reduced multiplet is called Ferrara-Zumino (FZ). When $\cY_\alpha$ can be set to zero, the reduced multiplet is called the $\cR$-multiplet. In this case the current $j_\mu$ is conserved, and it corresponds to a preserved $U(1)$ $R$-symmetry of the theory. The FZ-multiplet and the $\cR$-multiplet contain 12+12 degrees of freedom. If both $\chi_\alpha$ and $\cY_\alpha$ can be improved to 0 simultaneously, then the theory is superconformal and the corresponding multiplet is 8+8. 

Note that when $\chi_\alpha= 0$  also the string current (\ref{stringcur}) vanishes. Therefore, when the theory admits BPS strings, it is impossible to set $\chi_\alpha$ to zero by an improvement transformation, and the theory does not admit an FZ-multiplet. Analogously, if the theory admits BPS domain walls, then $\cY_\alpha$ cannot be improved to 0, and the theory does not admit an $\cR$-multiplet \cite{Dumitrescu:2011iu}.

\bibliographystyle{JHEP}
\bibliography{Bib_SUSY_boundary2}

\providecommand{\href}[2]{#2}\begingroup\raggedright\begin{thebibliography}{10}

\bibitem{Ooguri:1996ck}
H.~Ooguri, Y.~Oz, and Z.~Yin, {\it {D-branes on Calabi-Yau spaces and their
  mirrors}},  {\em Nucl.Phys.} {\bf B477} (1996) 407--430,
  [\href{http://xxx.lanl.gov/abs/hep-th/9606112}{{\tt hep-th/9606112}}].

\bibitem{Hori:2000ck}
K.~Hori, A.~Iqbal, and C.~Vafa, {\it {D-branes and mirror symmetry}},
  \href{http://xxx.lanl.gov/abs/hep-th/0005247}{{\tt hep-th/0005247}}.

\bibitem{Hori:2000ic}
K.~Hori, {\it {Linear models of supersymmetric D-branes}},
  \href{http://xxx.lanl.gov/abs/hep-th/0012179}{{\tt hep-th/0012179}}.

\bibitem{Govindarajan:1999js}
S.~Govindarajan, T.~Jayaraman, and T.~Sarkar, {\it {World sheet approaches to
  D-branes on supersymmetric cycles}},  {\em Nucl.Phys.} {\bf B580} (2000)
  519--547, [\href{http://xxx.lanl.gov/abs/hep-th/9907131}{{\tt
  hep-th/9907131}}].

\bibitem{Lindstrom:2002mc}
U.~Lindstrom, M.~Rocek, and P.~van Nieuwenhuizen, {\it {Consistent boundary
  conditions for open strings}},  {\em Nucl.Phys.} {\bf B662} (2003) 147--169,
  [\href{http://xxx.lanl.gov/abs/hep-th/0211266}{{\tt hep-th/0211266}}].

\bibitem{Albertsson:2002qc}
C.~Albertsson, U.~Lindstrom, and M.~Zabzine, {\it {N=1 supersymmetric sigma
  model with boundaries. 2}},  {\em Nucl.Phys.} {\bf B678} (2004) 295--316,
  [\href{http://xxx.lanl.gov/abs/hep-th/0202069}{{\tt hep-th/0202069}}].

\bibitem{Hassan:2003uq}
S.~Hassan, {\it {N=1 world sheet boundary couplings and covariance of
  nonAbelian world volume theory}},
  \href{http://xxx.lanl.gov/abs/hep-th/0308201}{{\tt hep-th/0308201}}.

\bibitem{Lindstrom:2002jb}
U.~Lindstrom and M.~Zabzine, {\it {N=2 boundary conditions for nonlinear sigma
  models and Landau-Ginzburg models}},  {\em JHEP} {\bf 0302} (2003) 006,
  [\href{http://xxx.lanl.gov/abs/hep-th/0209098}{{\tt hep-th/0209098}}].

\bibitem{Melnikov:2003zv}
I.~V. Melnikov, M.~R. Plesser, and S.~Rinke, {\it {Supersymmetric boundary
  conditions for the N=2 sigma model}},
  \href{http://xxx.lanl.gov/abs/hep-th/0309223}{{\tt hep-th/0309223}}.

\bibitem{Koerber:2003ef}
P.~Koerber, S.~Nevens, and A.~Sevrin, {\it {Supersymmetric nonlinear sigma
  models with boundaries revisited}},  {\em JHEP} {\bf 0311} (2003) 066,
  [\href{http://xxx.lanl.gov/abs/hep-th/0309229}{{\tt hep-th/0309229}}].

\bibitem{Herbst:2008jq}
M.~Herbst, K.~Hori, and D.~Page, {\it {Phases Of N=2 Theories In 1+1 Dimensions
  With Boundary}},  \href{http://xxx.lanl.gov/abs/0803.2045}{{\tt
  arXiv:0803.2045}}.

\bibitem{Horava:1995qa}
P.~Horava and E.~Witten, {\it {Heterotic and type I string dynamics from
  eleven-dimensions}},  {\em Nucl.Phys.} {\bf B460} (1996) 506--524,
  [\href{http://xxx.lanl.gov/abs/hep-th/9510209}{{\tt hep-th/9510209}}].

\bibitem{Altendorfer:2000rr}
R.~Altendorfer, J.~Bagger, and D.~Nemeschansky, {\it {Supersymmetric
  Randall-Sundrum scenario}},  {\em Phys.Rev.} {\bf D63} (2001) 125025,
  [\href{http://xxx.lanl.gov/abs/hep-th/0003117}{{\tt hep-th/0003117}}].

\bibitem{vanNieuwenhuizen:2005kg}
P.~van Nieuwenhuizen and D.~V. Vassilevich, {\it {Consistent boundary
  conditions for supergravity}},  {\em Class.Quant.Grav.} {\bf 22} (2005)
  5029--5051, [\href{http://xxx.lanl.gov/abs/hep-th/0507172}{{\tt
  hep-th/0507172}}].

\bibitem{Belyaev:2005rs}
D.~V. Belyaev, {\it {Boundary conditions in the Mirabelli and Peskin model}},
  {\em JHEP} {\bf 0601} (2006) 046,
  [\href{http://xxx.lanl.gov/abs/hep-th/0509171}{{\tt hep-th/0509171}}].

\bibitem{Belyaev:2005rt}
D.~V. Belyaev, {\it {Boundary conditions in supergravity on a manifold with
  boundary}},  {\em JHEP} {\bf 0601} (2006) 047,
  [\href{http://xxx.lanl.gov/abs/hep-th/0509172}{{\tt hep-th/0509172}}].

\bibitem{vanNieuwenhuizen:2006pz}
P.~van Nieuwenhuizen, A.~Rebhan, D.~V. Vassilevich, and R.~Wimmer, {\it
  {Boundary terms in supergravity and supersymmetry}},  {\em Int.J.Mod.Phys.}
  {\bf D15} (2006) 1643--1658,
  [\href{http://xxx.lanl.gov/abs/hep-th/0606075}{{\tt hep-th/0606075}}].

\bibitem{Andrianopoli:2014aqa}
L.~Andrianopoli and R.~D'Auria, {\it {N=1 and N=2 pure supergravities on a
  manifold with boundary}},  {\em JHEP} {\bf 08} (2014) 012,
  [\href{http://xxx.lanl.gov/abs/1405.2010}{{\tt arXiv:1405.2010}}].

\bibitem{Gaiotto:2008sa}
D.~Gaiotto and E.~Witten, {\it {Supersymmetric Boundary Conditions in N=4 Super
  Yang-Mills Theory}},  {\em J.Statist.Phys.} {\bf 135} (2009) 789--855,
  [\href{http://xxx.lanl.gov/abs/0804.2902}{{\tt arXiv:0804.2902}}].

\bibitem{Gaiotto:2008ak}
D.~Gaiotto and E.~Witten, {\it {S-Duality of Boundary Conditions In N=4 Super
  Yang-Mills Theory}},  {\em Adv.Theor.Math.Phys.} {\bf 13} (2009) 721,
  [\href{http://xxx.lanl.gov/abs/0807.3720}{{\tt arXiv:0807.3720}}].

\bibitem{Berman:2009kj}
D.~S. Berman and D.~C. Thompson, {\it {Membranes with a boundary}},  {\em
  Nucl.Phys.} {\bf B820} (2009) 503--533,
  [\href{http://xxx.lanl.gov/abs/0904.0241}{{\tt arXiv:0904.0241}}].

\bibitem{Berman:2009xd}
D.~S. Berman, M.~J. Perry, E.~Sezgin, and D.~C. Thompson, {\it {Boundary
  Conditions for Interacting Membranes}},  {\em JHEP} {\bf 1004} (2010) 025,
  [\href{http://xxx.lanl.gov/abs/0912.3504}{{\tt arXiv:0912.3504}}].

\bibitem{Okazaki:2013kaa}
T.~Okazaki and S.~Yamaguchi, {\it {Supersymmetric boundary conditions in
  three-dimensional N=2 theories}},  {\em Phys.Rev.} {\bf D87} (2013), no.~12
  125005, [\href{http://xxx.lanl.gov/abs/1302.6593}{{\tt arXiv:1302.6593}}].

\bibitem{Festuccia:2011ws}
G.~Festuccia and N.~Seiberg, {\it {Rigid Supersymmetric Theories in Curved
  Superspace}},  {\em JHEP} {\bf 1106} (2011) 114,
  [\href{http://xxx.lanl.gov/abs/1105.0689}{{\tt arXiv:1105.0689}}].

\bibitem{Dumitrescu:2012ha}
T.~T. Dumitrescu, G.~Festuccia, and N.~Seiberg, {\it {Exploring Curved
  Superspace}},  {\em JHEP} {\bf 1208} (2012) 141,
  [\href{http://xxx.lanl.gov/abs/1205.1115}{{\tt arXiv:1205.1115}}].

\bibitem{Klare:2012gn}
C.~Klare, A.~Tomasiello, and A.~Zaffaroni, {\it {Supersymmetry on Curved Spaces
  and Holography}},  {\em JHEP} {\bf 1208} (2012) 061,
  [\href{http://xxx.lanl.gov/abs/1205.1062}{{\tt arXiv:1205.1062}}].

\bibitem{Pasquetti:2011fj}
S.~Pasquetti, {\it {Factorisation of N = 2 Theories on the Squashed 3-Sphere}},
   {\em JHEP} {\bf 1204} (2012) 120,
  [\href{http://xxx.lanl.gov/abs/1111.6905}{{\tt arXiv:1111.6905}}].

\bibitem{Beem:2012mb}
C.~Beem, T.~Dimofte, and S.~Pasquetti, {\it {Holomorphic Blocks in Three
  Dimensions}},  {\em JHEP} {\bf 1412} (2014) 177,
  [\href{http://xxx.lanl.gov/abs/1211.1986}{{\tt arXiv:1211.1986}}].

\bibitem{Benini:2012ui}
F.~Benini and S.~Cremonesi, {\it {Partition functions of $\mathcal{N}=(2,2)$
  gauge theories on $S^2$ and vortices}},
  \href{http://xxx.lanl.gov/abs/1206.2356}{{\tt arXiv:1206.2356}}.

\bibitem{Bolle:1989ze}
D.~Bolle, P.~Dupont, and H.~Grosse, {\it {ON THE GENERAL STRUCTURE OF
  SUPERSYMMETRIC MODELS}},  {\em Nucl.Phys.} {\bf B338} (1990) 223--243.

\bibitem{Luckock:1991se}
H.~Luckock, {\it {Boundary terms for globally supersymmetric actions}},  {\em
  Int.J.Theor.Phys.} {\bf 36} (1997) 501--508.

\bibitem{Gates:1997kr}
S.~J. Gates, Jr., {\it {Ectoplasm has no topology: The Prelude}},  in {\em
  {Supersymmetries and quantum symmetries (SQS'97)}}, 1997.
\newblock \href{http://xxx.lanl.gov/abs/hep-th/9709104}{{\tt hep-th/9709104}}.

\bibitem{Gates:1998hy}
S.~J. Gates, Jr., {\it {Ectoplasm has no topology}},  {\em Nucl. Phys.} {\bf
  B541} (1999) 615--650, [\href{http://xxx.lanl.gov/abs/hep-th/9809056}{{\tt
  hep-th/9809056}}].

\bibitem{Howe:2011tm}
P.~S. Howe, T.~G. Pugh, K.~S. Stelle, and C.~Strickland-Constable, {\it
  {Ectoplasm with an Edge}},  {\em JHEP} {\bf 08} (2011) 081,
  [\href{http://xxx.lanl.gov/abs/1104.4387}{{\tt arXiv:1104.4387}}].

\bibitem{Belyaev:2008xk}
D.~V. Belyaev and P.~van Nieuwenhuizen, {\it {Rigid supersymmetry with
  boundaries}},  {\em JHEP} {\bf 0804} (2008) 008,
  [\href{http://xxx.lanl.gov/abs/0801.2377}{{\tt arXiv:0801.2377}}].

\bibitem{Bilal:2011gp}
A.~Bilal, {\it {Supersymmetric Boundaries and Junctions in Four Dimensions}},
  {\em JHEP} {\bf 1111} (2011) 046,
  [\href{http://xxx.lanl.gov/abs/1103.2280}{{\tt arXiv:1103.2280}}].

\bibitem{Komargodski:2010rb}
Z.~Komargodski and N.~Seiberg, {\it {Comments on Supercurrent Multiplets,
  Supersymmetric Field Theories and Supergravity}},  {\em JHEP} {\bf 1007}
  (2010) 017, [\href{http://xxx.lanl.gov/abs/1002.2228}{{\tt
  arXiv:1002.2228}}].

\bibitem{Dumitrescu:2011iu}
T.~T. Dumitrescu and N.~Seiberg, {\it {Supercurrents and Brane Currents in
  Diverse Dimensions}},  {\em JHEP} {\bf 1107} (2011) 095,
  [\href{http://xxx.lanl.gov/abs/1106.0031}{{\tt arXiv:1106.0031}}].

\bibitem{deAzcarraga:1989gm}
J.~de~Azcarraga, J.~P. Gauntlett, J.~Izquierdo, and P.~Townsend, {\it
  {Topological Extensions of the Supersymmetry Algebra for Extended Objects}},
  {\em Phys.Rev.Lett.} {\bf 63} (1989) 2443.

\bibitem{Hughes:1986dn}
J.~Hughes and J.~Polchinski, {\it {Partially Broken Global Supersymmetry and
  the Superstring}},  {\em Nucl.Phys.} {\bf B278} (1986) 147.

\bibitem{Gaiotto:2014kfa}
D.~Gaiotto, A.~Kapustin, N.~Seiberg, and B.~Willett, {\it {Generalized Global
  Symmetries}},  {\em JHEP} {\bf 02} (2015) 172,
  [\href{http://xxx.lanl.gov/abs/1412.5148}{{\tt arXiv:1412.5148}}].

\bibitem{Dierigl:2014xta}
M.~Dierigl and A.~Pritzel, {\it {Topological Model for Domain Walls in
  (Super-)Yang-Mills Theories}},  {\em Phys. Rev.} {\bf D90} (2014), no.~10
  105008, [\href{http://xxx.lanl.gov/abs/1405.4291}{{\tt arXiv:1405.4291}}].

\bibitem{Belyaev:2008ex}
D.~V. Belyaev and P.~van Nieuwenhuizen, {\it {Simple d=4 supergravity with a
  boundary}},  {\em JHEP} {\bf 09} (2008) 069,
  [\href{http://xxx.lanl.gov/abs/0806.4723}{{\tt arXiv:0806.4723}}].

\bibitem{Wess:1992cp}
J.~Wess and J.~Bagger, {\it {Supersymmetry and supergravity}}, .

\bibitem{Hori:2013ika}
K.~Hori and M.~Romo, {\it {Exact Results In Two-Dimensional (2,2)
  Supersymmetric Gauge Theories With Boundary}},
  \href{http://xxx.lanl.gov/abs/1308.2438}{{\tt arXiv:1308.2438}}.

\bibitem{Dvali:1996xe}
G.~Dvali and M.~A. Shifman, {\it {Domain walls in strongly coupled theories}},
  {\em Phys.Lett.} {\bf B396} (1997) 64--69,
  [\href{http://xxx.lanl.gov/abs/hep-th/9612128}{{\tt hep-th/9612128}}].

\bibitem{Chibisov:1997rc}
B.~Chibisov and M.~A. Shifman, {\it {BPS saturated walls in supersymmetric
  theories}},  {\em Phys.Rev.} {\bf D56} (1997) 7990--8013,
  [\href{http://xxx.lanl.gov/abs/hep-th/9706141}{{\tt hep-th/9706141}}].

\bibitem{Gorsky:1999hk}
A.~Gorsky and M.~A. Shifman, {\it {More on the tensorial central charges in N=1
  supersymmetric gauge theories (BPS wall junctions and strings)}},  {\em
  Phys.Rev.} {\bf D61} (2000) 085001,
  [\href{http://xxx.lanl.gov/abs/hep-th/9909015}{{\tt hep-th/9909015}}].

\bibitem{Gauntlett:2000ch}
J.~P. Gauntlett, G.~W. Gibbons, C.~M. Hull, and P.~K. Townsend, {\it {BPS
  states of D = 4 N=1 supersymmetry}},  {\em Commun.Math.Phys.} {\bf 216}
  (2001) 431--459, [\href{http://xxx.lanl.gov/abs/hep-th/0001024}{{\tt
  hep-th/0001024}}].

\bibitem{Shifman:2009zz}
M.~Shifman and A.~Yung, {\it {Supersymmetric solitons}}, .

\end{thebibliography}\endgroup

\end{document}